\documentclass[runningheads,openany]{svmult}

\usepackage{makeidx}   % allows index generation
\usepackage{graphicx}  % standard LaTeX graphics tool
\usepackage{subeqnar}  % subnumbers individual equations
\usepackage{multicol}  % used for the two-column index
\usepackage{physprbb}  % modified textarea for proceedings,
\makeindex             % used for the subject index
\begin{document}

%  this is Paris-00-intro
%
%
\def\AJ{Astron.~J. } 
\def\ARAA{Ann.~Rev.~Astron.~Astrophys. } 
\def\ApJ{Astrophys.~J. } 
\def\ApJL{Astrophys.~J.~Lett. } 
\def\ApJS{Astrophys.~J.~Supp. } 
\def\ApP{Astropart.~Phys. } 
\def\AA{Astron.~Astrophys. } 
\def\AAR{Astron.~Astrophys.~Rev. } 
\def\AAL{Astron.~Astrophys.~Lett. } 
\def\JGR{J.~Geophys.~Res. } 
\def\JPhG{J.~Phys.~G } 
\def\PhFl{Phys.~of~Fluids } 
\def\PR{Phys.~Rev. } 
\def\PRD{Phys.~Rev.~D } 
\def\PRL{Phys.~Rev.~Lett. } 
\def\Nature{Nature } 
\def\MNRAS{Mon.~Not.~R.~Astron.~Soc. } 
\def\ZA{Zeitschr. f{\"u}r Astrophys. } 
\def\ZFN{Zeitschr. f{\"u}r Naturforsch. } 
\def\etal{et al.}

\def\la{\mathrel{\mathpalette\fun <}}
\def\ga{\mathrel{\mathpalette\fun >}}
\def\fun#1#2{\lower3.6pt\vbox{\baselineskip0pt\lineskip.9pt
  \ialign{$\mathsurround=0pt#1\hfil##\hfil$\crcr#2\crcr\sim\crcr}}}
\def\simle{\lower 2pt \hbox {$\buildrel < \over {\scriptstyle \sim }$}}
\def\simge{\lower 2pt \hbox {$\buildrel > \over {\scriptstyle \sim }$}}
\newif\iffigures
%______________________________________________________________________
%% choose ONE of the following options by omitting %% at the beginning
%______________________________________________________________________
%      \figurestrue  %% <<<<--- figures are included %%
        \figuresfalse %% <<<<--- picplace box used instead of figures

\title*{Introduction to Cosmic Rays\thanks{introductory
chapter to "Physics and Astrophysics of Ultra-High-Energy Cosmic Rays",
Lecture Notes in Physics vol.~576 (eds.: M.Lemoine, G.Sigl), based
on UHECR2000 (Meudon, June 26-29, 2000), for online version see
http://link.springer.de/link/service/series/2669/tocs/t1576.htm,
copyright Springer-Verlag Berlin Heidelberg 2001.}}

\titlerunning{Introduction to Cosmic Rays}

\author{Peter~L.~Biermann\inst{1}
\and G\"unter Sigl\inst{2}}

\authorrunning{Peter~L.~Biermann and G\"unter~Sigl}

\institute{Max-Planck Institute for Radioastronomy, and\\
Department for Physics and Astronomy, University of Bonn,
Bonn, Germany
\and Institut d'Astrophysique de Paris, Paris, France}

%  
%  version May 13, 2001 -- sent from GS to PLB May 27, 2001
%  with further corrections June 2 by PLB
%  and some more corrections June 24, 2001 by PLB
%  sigl@iap.fr
%

\maketitle

\begin{abstract}
Energetic particles, traditionally called {\it Cosmic Rays},
were discovered nearly a hundred years ago, and their origin is still
uncertain.  Their main constituents are the normal nuclei as in the
standard cosmic abundances of matter, with some enhancements for the
heavier elements; there are also electrons, positrons and anti-protons.  
Today we also have information on isotopic abundances, which show some
anomalies, as compared with the interstellar medium.  And there is
antimatter, but no anti-nuclei.  The known spectrum extends over energies
\index{cosmic ray!spectrum}
from a few hundred MeV to 300 EeV ($=3\times10^{20}$\,eV), and shows
few clear spectral signatures:  There is a small spectral break near $5
\times 10^{15}$ eV, commonly referred to as the {\it knee}, where the
\index{knee, cosmic ray spectrum}
spectrum turns down; there is another spectral break near $3
\times 10^{18}$ eV, usually called the {\it ankle}, where the spectrum
\index{ankle, cosmic ray spectrum}
turns up again.  Up to the ankle the cosmic rays are usually interpreted
as originating from supernova\index{supernova} explosions, i.e.
those cosmic ray particles
are thought to be Galactic in origin; however, the details are not
clear.  We do not know what the origin of the knee is, and what physical
processes can give rise to particle energies in the energy range from the
knee to the ankle.  The particles beyond the ankle have to be
extragalactic, it is usually assumed, because the Larmor radii in the
\index{Larmor!radius}
Galactic magnetic field are too large; this argument could be overcome if
\index{magnetic field!galactic}
those particles were very heavy nuclei as Fe, an idea which appears to be
\index{heavy nucleus!primary}
inconsistent, however, with the airshower data immediately above the
energy of the ankle.  Due to interaction with the cosmic microwave
background (CMB),
\index{cosmic microwave background (CMB)}
a relic of the Big Bang, there is a strong cut-off expected near 50 EeV
(=$5 \times 10^{19}$ eV), which is,  however, not seen; this expected
cutoff is called the GZK-cutoff  after its discoverers, Greisen, Zatsepin
\index{Greisen-Zatsepin-Kuzmin effect (GZK)}
and Kuzmin.  The spectral index
$\alpha$ is near 2.7 below the knee, near 3.1 above the knee, and again
near 2.7 above the ankle, where this refers to a differential spectrum
of the form $E^{-\alpha}$ in numbers.  The high energy cosmic rays
beyond the GZK-cutoff are the challenge to interpret.  We will describe
the various approaches to understand the origin and physics of cosmic
rays.
\end{abstract}

\section{Introduction and History}

Cosmic Rays were discovered by Hess~\cite{1:CRA} and
Kohlh{\"o}rster~\cite{1:CRB} in the beginning
of the twentieth century through their ionizing effect on airtight vessels
of glas enclosing two electrodes with a high voltage between them.  This
ionizing effect increased with altitude during balloon flights, and
therefore the effect must come from outside the Earth.  So the term {\it
Cosmic Rays} was coined.  The Earth's magnetic field acts on energetic
\index{geomagnetic field!}
particles according to their charge, they are differently affected
coming from East and West, and so their charge was detected, proving
once and for all that they are charged particles.  At the 
energies near $10^{18}$ eV there is observational evidence, that a small
fraction of the particles are neutral, and in fact neutrons; these
events correlate on the sky with the regions of highest expected cosmic
ray interactions, the Cygnus region and the Galactic center region.  From
around 1960 onwards particles were detected at or above
$10^{20}$ eV, with today about two dozen such events known.  It took
almost forty years for the community to be convinced that these energies
are real, and this success is due to the combination of air fluorescence
data with ground-based observations of secondary electrons/positrons and
muons, as well as \v{C}erenkov light; the Fly's Eye~\cite{1:fe},
\index{air shower, extensive (EAS)!muon component}
\index{Fly's Eye experiment}
\index{Cerenkov@\v{C}erenkov!light}
Haverah Park~\cite{1:haverah} and AGASA~\cite{1:agasa}
\index{Haverah Park ground array}
\index{Akeno Giant Air Shower Array (AGASA)}
arrays are those with the most extensive discussion of their data
out and published; other arrays have also contributed a great deal, like
Yakutsk~\cite{1:yakutsk}, Volcano Ranch~\cite{1:volcano} and 
\index{Yakutsk ground array}
\index{Volcano Ranch ground array}
SUGAR~\cite{1:sugar}.  Already in the fifties it was noted that
\index{SUGAR ground array}
protons with energies above $3\times10^{18}$ eV have Larmor radii in
\index{Larmor!radius}
the Galactic magnetic field which are too large to be contained, and
\index{magnetic field!galactic}
so such particles must come from outside \cite{1:Cocconi56}.  After the
CMB was discovered, in the early 1960s, it was
\index{cosmic microwave background (CMB)}
noted only a little later by Greisen~\cite{1:greisen}, and Zatsepin
\index{Greisen-Zatsepin-Kuzmin effect (GZK)}
and Kuzmin~\cite{1:zat-kuz}, in two papers, that near and above an energy of
$5 \times 10^{19}$ eV (called the GZK-cutoff) the interaction with the
CMB would lead to strong losses, if these particles were
\index{cosmic ray!energy loss}
protons, as is now believed on the basis of detailed airshower data.  In
such an interaction, protons see the photon as having an energy of above
the pion mass, and so pions can be produced in the reference frame of the
\index{pion production}
collision, leading to about a 20 \% energy loss of the proton about
\index{interaction length}
every $\simeq6\,$Mpc in the
observer frame.  Therefore for an assumed cosmologically homogeneous
distribution of sources for protons at extreme energies, a spectrum at
\index{cosmic ray!spectrum}
Earth is predicted which shows a strong cutoff at  $5 \times 10^{19}$ eV,
the GZK-cutoff.  This cutoff is not seen, leading to many speculations as
\index{Greisen-Zatsepin-Kuzmin effect (GZK)}
to what the nature of the particles beyond the GZK-energy, and their
origin might be.

Cosmic rays are measured with balloon flights, satellites, now
with instruments such as AMS~\cite{1:AMS} on the Space Shuttle, and
soon also with instruments on the International Space Station~\cite{1:iss},
and with Ground Arrays. The instrument
chosen depends strongly on what is being looked for, and the energy of
the primary particle.  One of the most successful campaigns has been with
balloon flights in Antarctica, where the balloon can float at about 40 km
altitude and circumnavigate the South Pole once, and possibly even
several times during one Antarctic summer.  For very high precision
measurements very large instruments on the Space Shuttle or soon the
International Space Station have been or will be used, such as for the
search for antimatter.  The presently developed new experiments
\index{antimatter}
such as the fluorescence detector array HiRes~\cite{1:hires} and the
\index{High Resolution Fly's Eye (HiRes)}
hybrid array Auger~\cite{1:auger}, are expected to contribute decisively
\index{Pierre Auger Observatory}
\index{hybrid experiment}
to the next generation of data sets for the highest energies.

Critical measurements are today the exact spectrum of the most common
elements, Hydrogen and Helium, the energy dependence of the fraction of
anti-particles (anti-protons and positrons), isotopic ratios of elements
\index{antimatter}
such as Neon and Iron, the ratio of spallation products such as Boron to
the primary nuclei such as Carbon as a function of energy, the chemical
\index{heavy nucleus!primary}
\index{cosmic ray!composition}
composition near and beyond the knee, at about $5 \times 10^{15}$ eV,
\index{knee, cosmic ray spectrum}
and the spectrum and nature of the particles beyond the ankle, at
\index{cosmic ray!spectrum}
\index{ankle, cosmic ray spectrum}
$3 \times 10^{18}$ eV, with special emphasis on the particles beyond the
expected GZK-cutoff, at $\simeq5\times 10^{19}$ eV.  The detection of
\index{Greisen-Zatsepin-Kuzmin effect (GZK)}
anti-nuclei would constitute a rather extreme challenge. One of the most
decisive points is the quest for the highest energy events and the high
energy cutoff in the spectrum. This is also the main topic of
the present volume. The data situation and experimental issues
involved at the highest energies have been reviewed in
Refs.~\cite{1:data,1:NW2000}.

Relevant reviews and important original papers have been published over
many years, e.g.,
\cite{1:Fermi,1:G53a,1:G53b,1:GS63,1:Gaisser90,1:G93,1:G96,1:Venyabook}.

\section{Physical Concepts}

\subsection{Cosmic Ray Spectrum and Isotropy}\label{1:s1}

\index{cosmic ray!spectrum|(}
\begin{figure}%[t]
\begin{center}
\includegraphics[width=.98\textwidth,clip=true]{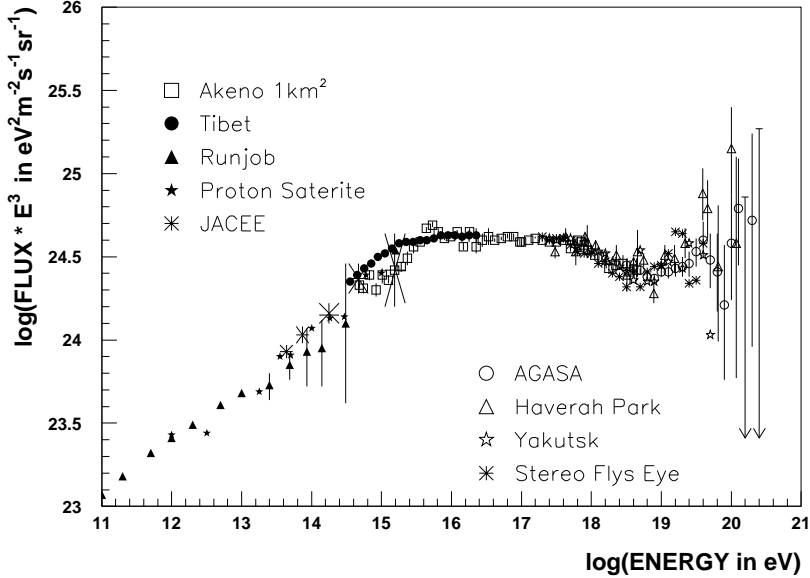}
\end{center}
\caption[...]{The CR all-particle spectrum observed
by different experiments above $10^{11}\,$eV (from
Ref.~\cite{1:NW2000}).  The differential flux in units of events per
\index{cosmic ray!flux}
area, time, energy, and solid angle was multiplied with $E^3$ to
project out the steeply falling character. The ``knee'' can be seen at
\index{knee, cosmic ray spectrum}
$E\simeq4\times 10^{15}\,$eV, the ``second knee'' at
$\simeq3\times10^{17}\,$eV, and the ``ankle'' at
\index{ankle, cosmic ray spectrum}
$E\simeq5\times10^{18}\,$eV}
\label{1:F1}
\end{figure}

The number of particles at a certain energy $E$ within a certain small
energy interval $d E$ is called the spectrum.  Cosmic rays have
usually a powerlaw spectrum, which is referred to as a non-thermal
behaviour, since non-thermal processes are thought to be producing
such spectra. Flux is usually expressed as the number of particles,
coming in per area, per second, per solid angle in steradians (all sky
is 4 $\pi$), and per energy interval. Cosmic rays have a spectrum near
$E^{-2.7}$ up the the knee, at about $5 \times 10^{15}$ eV, and then
\index{knee, cosmic ray spectrum}
about $E^{-3.1}$ beyond, up the ankle, at about $3 \times 10^{18}$ eV,
\index{ankle, cosmic ray spectrum}
beyond which the spectrum becomes hard to quantify, but can very
approximately again be described by $E^{-2.7}$. There is no other
strong feature in the spectrum, especially no cutoff at the upper end.
There is some limited evidence from the newest experiments
(AGASA~\cite{1:agasa} and HiRes~\cite{1:hires,1:Hires2000}) for
\index{High Resolution Fly's Eye (HiRes)}
\index{Akeno Giant Air Shower Array (AGASA)}
another feature, at about $3 \times 10^{17}$ eV, called {\it the
second knee}, where the spectrum appears to dip.  Both the first and
the second knee may be at an energy which is proportional to charge
\cite{1:Peters}, i.e. at a constant Larmor radius, and therefore
\index{Larmor!radius}
may imply a range in energies per particle. Figure~\ref{1:F1}
shows the overall cosmic ray spectrum.

There is no anisotropy except for a weak hint near
\index{arrival direction distribution!isotropy}
$10^{18}\,$eV~\cite{1:fe-aniso,1:agasa-aniso,1:clayetal}, and the suggestive
signal for pairing at energies near and beyond the
GZK-cutoff~\cite{1:clustering}.
\index{Greisen-Zatsepin-Kuzmin effect (GZK)}
\index{clustering!ultra-high energy cosmic ray arrival direction}
\index{arrival direction distribution!anisotropy}
\index{cosmic ray!spectrum|)}

\subsection{Fermi Acceleration}\index{acceleration|(}
\index{acceleration!diffusive shock|(}

  In a compressing system the particles gain energy; the walls can be
magnetic irregularities which reflect charged particles through
\index{magnetic field!inhomogeneous}
\index{magnetic field!wave}
magnetic resonance between the gyromotion and waves in the ionized
magnetic gas, the plasma.  Such magnetic irregularities usually exist
everywhere in a plasma that gets stirred by, e.g., stellar ultraviolet
radiation and their ionization fronts, by stellar winds, supernova
\index{supernova} explosions, and by the energetic particles moving through.
Considering now the two sides of a shock, one realizes that this is a
permanently compressing system for charged particles which move much
faster than the flow in the shock frame.  Therefore particles gain
energy, going back and forth.  In one cycle they normally gain a
fraction of $U_{sh}/c$ in momentum (adopting relativistic particles
here), and the population loses a fraction of also $U_{sh}/c$.  Here
$U_{sh}$ is the shock velocity. For the original articles by E.~Fermi
\index{shock!velocity}
see Ref.~\cite{1:Fermi}, Ref.~\cite{1:BE87} for a recent review, and
see also the contribution by G.~Pelletier in this volume.

The density jump $r$ in an adiabatic shockfront is given by the
\index{shock!adiabatic}
adiabatic index of the gas $\gamma$ and the upstream Mach number of
the shock $M_1$

\begin{equation}
r \, = \, \frac{\gamma +1}{\gamma -1 +2/M_1^2}
\end{equation}

The general expression for the spectral index of the
particle momentum distribution $p^{-a}$ is

\begin{equation}
a \, = \, \frac{3 r}{r-1} 
\end{equation}

This is in three-dimensional phase space; the energy distribution then is
given by $E^{2-a}$, for relativistic particles.   This means, for
instance, that for a very large Machnumber and the standard case of
$\gamma = 5/3$ the density jump is 4, and the spectral index is $a - 2 =
2$.  For $\gamma = 4/3$, as would be the case in a gas with a relativistic
equation of state (like a radiation dominated gas) the density jump is 7,
and the spectral energy index of the particles is $a - 2 = 3/2$.
The time scale for acceleration is given in, e.g.,
\cite{1:Drury83,1:Jokipii87}. 

In a relativistic shock wave the derivation no longer holds so simply for
\index{shock!relativistic}
the spectrum; however, it is worth noting that the density jump can go to
infinity both in the case of a relativistic shockwave as in the case of a
strong cooling shock.  Then the spectral index in energy approaches $a -
2 = 1$.  However, detailed Monte-Carlo simulations for relativistic
shocks, taking into account the highly anisotropic nature of the
scattering as well as the particle distribution, again find a spectrum
near 2~\cite{1:BedO98}. For more details on Fermi acceleration see also the
contribution by G.~Pelletier in this volume.
\index{acceleration!diffusive shock|)}
\index{acceleration|)}

\subsection{Spallation}  

Spallation is the destruction of atomic nuclei in a
         collision with another energetic particle, such as another
         nucleus, commonly a proton
\cite{1:GM77,1:GM87,1:Tsao2001}.  In this destruction many pieces of
         debris can be formed, with one common result the stripping of
         just one proton or neutron, and another common result a
         distribution of lighter nuclei.  Since the proton number
         determines the chemical element, these debris are usually other
         nuclei, such as Boron, from the destruction of a Carbon nucleus. 
It is an interesting question, whether these collisions lead to a new
state of matter, the quark-gluon plasma; the Relativistic Heavy Ion
\index{quark}
Collider (RHIC) experiment~\cite{1:rhic} performed at
Brookhaven collides heavy nuclei with each other, in order to find
evidence for this new state.  Both in our upper atmosphere and out in
\index{atmosphere}
the Galaxy such collisions happen all the time, at very much higher
energy than possible in the laboratory, and may well be visible in the
data.  Conversely, the existing data could be used perhaps to derive
limits on what happens when a quark-gluon plasma is formed.

As a curiosity we mention that collisions of energetic cosmic rays
with each other and with large objects such as the moon have been
used to constrain the risk that high energy collisons in terrestrial
accelerators could produce particles or new vacuum states that would
trigger a phase transition to a lower energy state such as strange
quark matter which would destroy the Earth~\cite{1:risks}. This risk can
\index{quark}
be determined by calculating how much more often such processes
occurred naturally involving cosmic rays since the birth of our
Universe.

\subsection{Chemical Abundances}\label{1:s2}
\index{cosmic ray!composition|(}
The chemical abundances in cosmic rays are rather similar to first
approximation to those in the interstellar medium~\cite{1:Reeves}.
We consider them in the following framework:
We plot the number of particles per energy interval as a function of
energy per particle, and normalize at 1 TeV energy per particle, so as to
be free of any solar modulation effect~\cite{1:LB-CR}. And we refer to
\index{solar!modulation}
Silicon for the comparison, so by definition the abundance for Silicon is
adopted to be equal for cosmic rays and for the so-called cosmic
abundances in the interstellar medium.  In this well defined frame-work we
then note the following differences:

\begin{itemize}

\item{}  The abundance of Hydrogen is very much less for cosmic rays, as
is the ratio of Hydrogen to Helium.

\item{}  The abundances of the elements Lithum, Beryllium and Boron are
very much larger in cosmic rays than in the interstellar medium, by
several powers of ten.  

\item{}  The abundances of the sub-Iron elements are also larger than
relative to Iron for cosmic rays.

\item{}  The abundances of odd-Z elements are larger.

\item{}  And, finally, those elements with a low first ionization
potential are systematically more abundant.

\end{itemize}

\begin{figure}%[t]
\begin{center}
\includegraphics[width=.98\textwidth,clip=true]{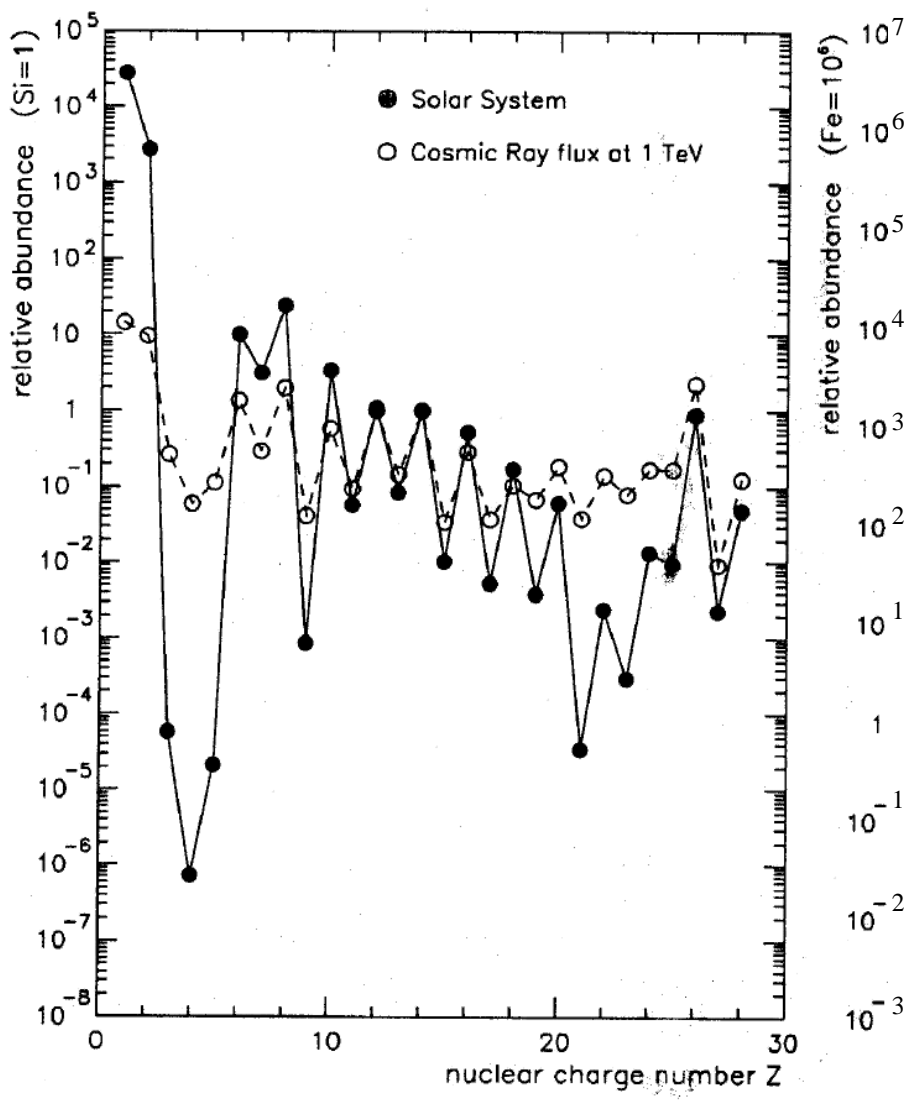}
\end{center}
\caption[...]{The chemical composition of cosmic rays relative
to Silicon and iron at 1 TeV, and in the solar System, as a function
of nuclear charge $Z$, from Ref.~\cite{1:LB-CR}}
\label{1:F2}
\end{figure}

These tendencies can be seen in Fig.~\ref{1:F2} which compares
solar System abundances with abundances in cosmic rays at
1 TeV.

In addition, the isotopic ratios among a given element are sometimes
very similar to those in the interstellar medium, and for other cases,
very different, indicating rather specific source contributors.

In all versions of theories it is acknowledged that spallation of 
abundant elements plays a major role, especially for the light elements,
where spallation and subsequent ionization loss can even explain the
\index{cosmic ray!energy loss}
abundances of the light elements in the interstellar medium.  This is an
especially interesting test using the light element abundances in stars
formed in the young years of our Galaxy~\cite{1:R97a}.
\index{cosmic ray!composition|)}

\subsection{Cosmic Ray Airshower}  \index{air shower, extensive (EAS)|(}

When a primary particle at high energy, either a photon, or a nucleus,
         comes into the upper atmosphere, the sequence of interactions
\index{atmosphere}
         and cascades form an airshower.  This airshower can be
\index{cascade!atmospheric}
         dominated by \v{C}erenkov light, a bluish light, produced when
\index{Cerenkov@\v{C}erenkov!light}
         particles travel at a speed higher than the speed of light
         $c$ divided by the local index of refraction (which is 4/3 in
         water, for instance, and about 1.0003 in air).  Observing
         this bluish light allows observations of high GeV to TeV
         photon sources in the sky.  For particles, such as protons,
         or atomic nuclei, the resulting airshower is dominated by air
\index{air shower, extensive (EAS)!heavy nucleus induced}
         fluorescence, when normal emission lines of air molecules are
\index{fluorescence!light}
         excited, and by a pancake of secondary electrons and
         positrons as well as muons.  Most modern observations of very
\index{air shower, extensive (EAS)!muon component}
         high energy cosmic rays are done either by observing the air
         fluorescence, (arrays such as Flys's Eye~\cite{1:fe},
         HiRes~\cite{1:hires}, or Auger~\cite{1:auger}), or by
         observing the secondary electrons and positrons (in arrays
         such as Haverah Park~\cite{1:haverah}, AGASA~\cite{1:agasa},
\index{Akeno Giant Air Shower Array (AGASA)}
\index{Yakutsk ground array}
\index{Pierre Auger Observatory}
\index{Haverah Park ground array}
\index{Fly's Eye experiment}
\index{High Resolution Fly's Eye (HiRes)}
         Yakutsk~\cite{1:yakutsk}, or also Auger~\cite{1:auger}). In
         the further future such observations may be possible from
         space, by observing the air fluorescence, or also the
         reflected \v{C}erenkov light, from either the International
\index{Cerenkov@\v{C}erenkov!light}
         Space Station, or from dedicated satellites.  Fly's Eye was
         and HiRes is in Utah, USA, Auger is in Argentina, AGASA is in
         Japan, Yakutsk is in Russia, and Haverah Park was in the
         United Kingdom.  Future planned experiments are
         EUSO~\cite{1:euso} on the space station, built by the ESA,
\index{EUSO (Extreme Universe Space Observatory)}
         and, even later, a satellite mission, OWL~\cite{1:owl},
         \index{OWL (Orbiting Wide-angle Light collector)} discussed
         by NASA. For reviews of experimental techniques to
         detect giant airshowers see Refs.~\cite{1:data,1:NW2000}.
\index{air shower, extensive (EAS)|)}

\subsection{Cosmic Ray GZK-Cutoff}  

The interactions with the CMB should produce
\index{Greisen-Zatsepin-Kuzmin effect (GZK)}
\index{cosmic microwave background (CMB)}
a strong cutoff in the observed spectrum, at $5 \times 10^{19}$ eV,
called the GZK-cutoff~\cite{1:greisen,1:zat-kuz,1:UHECR}. This
\index{Greisen-Zatsepin-Kuzmin effect (GZK)}
is expected provided that a) these particles are protons (or neutrons), and
b) the source distribution is homogeneous in the universe.  This
cutoff is not seen; in fact, no cutoff is seen at any energy, up
to the limit of data, at $\simeq3\times 10^{20}$ eV, or 300 EeV.  This
is one of the most serious problems facing cosmic ray physics
today.  Assuming a source distribution just as the observed
galaxy distribution alleviates the problem, but does not solve
it~\cite{1:GMT99b,1:BBO2001} (see also the contribution by G.~Medina Tanco
in this volume).

\subsection{Black Holes} 
 
%Compressing a star to a miniscule size, in case of the Sun to a radius
%of 3 km, makes it impossible for any radiation to come out anymore;
%all particles and radiation hitting such an object disappear from this
%world.  This is called a black hole.  
\index{black hole}\index{black hole!supermassive}
It is now believed that almost all galaxies have a massive black hole
at their center, with masses sometimes ranging up $10^{10}$ solar
masses, but usually much less.  There are also stellar mass black
holes, but their number is not well known, probably many thousands in
each galaxy.  The growth of these black holes has almost certainly put
an enormous amount of energy into the universe, possibly commensurate
with other forms of baryonic energy.  The ratio of the masses of the
black holes and the stellar spheroidal component of older stars has a
narrow distribution which is limited from above by about 1:300.  There
is a near perfect correlation between black hole mass and the velocity
dispersion of the inner stars of the central cusp around the black
\index{central region}
hole~\cite{1:KR95,1:Faber97,1:Magorrian98,1:Gebhardt2000}.  
These black holes can be expected to interact strongly with their
environment, both in stars and in gas \cite{1:LSh78}.

\subsection{Our Galaxy}  

Our galaxy is a flat distribution of stars and gas, mixed with
interstellar dust, and embedded in a spheroidal distribution of old
stars.  The age of this system is about 15 billion %(=$15 \times 10^9$)
years; its size is about 30 kpc across, and its inner region is about
6 kpc across.  At its very center there is a black hole with $2.6
\index{black hole!supermassive}
\times 10^6$ solar masses~\cite{1:MF2001}. 
         The gravitational field is dominated in the outer parts of the
         Galaxy by an unknown component, called dark matter, which we
\index{dark matter}
         deduce only through its gravitational force.  In the innermost
         part of the galaxy normal matter dominates.  The mass ratio of
         dark matter to stars to interstellar matter in our Galaxy is
         about 100:10:1.  Averaged over the nearby universe these ratios
         are shifted in favor of gas, with gas dominating over stars
         probably, but with dark matter still dominating over stars and
         gas by a large factor.  The universal ratios of baryonic
matter, dark matter and the $\Lambda$-term have been tightly constrained
by the observation of the first three waves in the fluctuation spectrum
of the CMB by the balloon experiments
\index{cosmic microwave background (CMB)}
BOOMERanG~\cite{1:boomerang}, MAXIMA~\cite{1:maxima}, and the
ground detector DASI~\cite{1:dasi}, as well as by measurements of the
relation between apparent magnitude and redshift of certain type Ia
supernovae\index{supernova} which serve as ``standard candles'' of
known absolute
luminosity~\cite{1:typeIa}. All experiments agree rather well in
these conclusions~\cite{1:combined}. In very small galaxies the dark
matter component dominates over baryonic matter even at the center
\cite{1:Ostriker93,1:Mateo98}.
\index{central region}

\subsection{Interstellar Matter}  

The gas in between the stars in our Galaxy is composed of very hot
gas (order  $4 \times 10^6$ K), various stages of cooler gas, down
to about 20 K, dust, cosmic rays, and magnetic
\index{magnetic field!galactic}
fields~\cite{1:ROSAT97,1:Kaneda97,1:Valinia98,1:Pietz98}.
All three components, gas, cosmic
         rays, and magnetic fields, have approximately the same energy
         density, which happens to be also close to the energy density of
         the CMB, about 1 eV per cm$^3$.  The average
\index{cosmic microwave background (CMB)}
         density of the neutral hydrogen gas, of temperature a few $10^3$
         K, is about 1 particle per cm$^3$, in a disk of thickness about
         100 pc (= $3 \times 10^{20}$ cm).  The very hot gas extends much
         farther from the symmetry plane, about 2 kpc on either side.

\subsection{Magnetic Fields}
\index{magnetic field|(}

Magnetic fields are ubiquitous in the
Universe~\cite{1:Kronberg94,1:Beck96,1:Vallee97} (see also the
contribution by G.~Medina~Tanco in this volume).  In our Galaxy they
have a total strength of about 6 - 7 microGauss ($\mu$G) in the solar
neighborhood, and about 10$\,\mu$G further in, at around 3 kpc from
the center.  The magnetic field is partially irregular, partially
regular, with roughly 1/2 to 2/3 of it in a circular ring-like
pattern; other galaxies demonstrate that the underlying symmetry is
dominated by a spiral structure with the overall magnetic field
pointing inwards along the spiral.  One level down in scale, the fine
structure is then of occasional reversals, but still mostly parallel
to a circle around the center.  At the small scales, less than the
thickness of the hot disk, it appears that the magnetic field can be
described as a Kolmogorov turbulence spectrum~\cite{1:kolmogorov}, all
\index{magnetic field!turbulence spectrum}
the way down to dissipation scales.
\index{magnetic field!galactic}
\index{magnetic field!inhomogeneous}

The origin of the magnetic field is not understood
\index{magnetic field!origin}
\cite{1:Krause98,1:Kulsrud99,1:GR2001}.  Comparing our Galaxy with
others, in the starburst phase, and also at high redshift makes it
obvious that the magnetic field is regenerated at time scales which
are less or at most equal to the rotation time scale, with
circumstantial evidence suggesting that this happens at a few times
$10^7$ years.  Interestingly, this is the same time scale at which
convection losses transport energy from the disk of the Galaxy, and on
\index{cosmic ray!energy loss}
which cosmic ray energy is lost. We do not have a real understanding
of what drives the energy balance of the interstellar medium.
\index{magnetic field|)}

\subsection{Transport of Cosmic Rays}
\index{cosmic ray!transport|(}
From the ratio of radioactive isotopes resulting from spallation to
stable isotopes we can deduce the time of transport of cosmic rays near 1
GeV:  It is about $3 \times 10^7$ years. This is very similar to the sound
crossing time scale across the hot thick disk of the interstellar medium,
and also to the Alfv{\'e}nic time scale across the same thick disk.  It is
unlikely that these numerical coincidences are chance.

The transport of cosmic rays is dominated by a variety of effects
\cite{1:GM77,1:GM87,1:Tsao2001}:

\begin{itemize}

\item{}  Ionization losses, mosty relevant for protons and nuclei.  This
\index{heavy nucleus!energy loss}
\index{cosmic ray!energy loss}
limits the lower energy of protons to about 50 MeV {\it after} traversing
most of the interstellar medium path, as derived from the ionizing effect
\cite{1:Nath94}.

\item{}  Spallation - discussed separately above.  For any given isotope,
spallation is a loss and a gain-process in the equation of balance.
\index{cosmic ray!energy loss}

\item{}  Radioactive decay.  For any specific isotope this can be a loss
\index{cosmic ray!energy loss}
and a gain-process.  The resulting observed ratios provide a clock for
\index{cosmic ray!clock}
cosmic ray transport.

\item{}  Synchrotron and Inverse Compton losses, only relevant for
\index{synchrotron!energy loss}
\index{inverse Compton scattering (ICS)}
\index{cosmic ray!energy loss}
electrons and positrons.  Above about 10 GeV these losses dominate over
diffusive losses, and so the spectrum is steepened by unity.  Then one
deduces from the observed spectrum of $E^{-3.3}$, that injection must have
\index{cosmic ray!spectrum}
happened with about $E^{-2.3}$.
\index{injection}

\item{}  Diffusive loss from the disk.  This is almost certainly governed
\index{cosmic ray!escape}
by the spectrum of turbulence, in an isotropic approximation best
\index{magnetic field!turbulence spectrum}
described by a Kolmogorov spectrum~\cite{1:kolmogorov}.
This entails that the time scale of loss is proportional to $E^{-1/3}$. 
In an equilibrium situation this steepens the observed spectrum by 1/3
over the injection spectrum.  Along this line of reasoning one deduces,
\index{injection}
that without re-acceleration the injection spectrum ought to be
\index{re-acceleration, cosmic rays}
$E^{-2.35}$ approximately, as noted immediately above providing a very
important consistency check.
\index{cosmic ray!spectrum}

\item{}  Convective loss from the disk.  This is likely to dominate at
\index{cosmic ray!escape}
energies below about 1 GeV for protons, or the corresponding energy of
other nuclei with the same Larmor radius.
\index{Larmor!radius}

\item{}  Magnetic field irregularities; in analogy with the Sun, it is
\index{magnetic field!inhomogeneous}
conceivable that the magnetic field is very inhomogeneous, contains
flux tubes of much higher than average field, and then the transport of
cosmic ray particles is governed by a mixture of streaming, convection,
and diffusion by pitch angle scattering on these magnetic irregularities.
\index{pitch angle!scattering}
\index{diffusion}

\item{}  Some cosmic rays almost certainly come from outside the Galaxy,
coming down the galactic wind - of which the existence is very likely, but
\index{galactic!wind}
not certain.  Using then the analogy with the solar wind, we need to
\index{solar!wind}
again ask the question what the most likely turbulence spectrum is in
\index{magnetic field!turbulence spectrum}
the wind, and that may be quite different from a Kolmogorov
spectrum~\cite{1:kolmogorov},
$k^{-5/3}$, where $k$ is the wavenumber, and the spectrum denotes the
energy per volume per wavenumber in isotropic phase space. Such a
Kolmogorov spectrum is observed in the solar wind over some part of the 
wavenumber spectrum.  Some have argued that it could be governed by the
repeated injection of supernova shockwaves, and so best be described by a
\index{supernova!shock wave}
$k^{-2}$ spectrum.  Interestingly, for just such a spectrum the
scattering in the irregularities of cosmic ray particles becomes
independent of energy, and so there would be no critical energy, below
which the cosmic ray spectrum coming in from the outside is cut off. 
This situation would then be quite different from the solar wind, where
all cosmic rays below about 500 MeV/nucleon (measured on the outside) are
cut off altogether.

\end{itemize}

The transport of ultra-high energy cosmic rays, photons, and
neutrinos in extragalactic space is dominated by various
\index{neutrino!propagation}
processes: Pion production (leading to the GZK effect) for
\index{pion production}
\index{Greisen-Zatsepin-Kuzmin effect (GZK)}
nucleons above $\simeq 5\times10^{19}\,$eV, electromagnetic
\index{nucleon!propagation}\index{nucleon!interaction}
\index{nucleon!energy loss}
cascades for $\gamma-$rays and, under certain circumstances,
\index{cascade!electromagnetic}
weak interactions with, for example, production and decay
of Z-bosons for ultra-high energy neutrinos propagating from
\index{Z-burst scenario}
large redshifts. Furthermore, protons and nuclei are significantly
\index{deflection, magnetic}
deflected by or even diffuse in large scale extragalactic
\index{diffusion}
\index{deflection, magnetic}
magnetic fields \cite{1:RKB98}. For a detailed discussion of these effects
\index{magnetic field!extragalactic (EGMF)}
see the contribution by G.~Sigl in this volume.
\index{cosmic ray!transport|)}

\subsection{Supernovae}  
\index{supernova}
All stars above an original mass of more than 8
         solar masses are expected to explode at the end of their
         life-time, after they have exhausted nuclear burning; the
         observable effect of such an explosion is called a supernova. 
         When they explode, they emit about $3 \times 10^{53}$ erg in                   neutrinos,\index{supernova!neutrino production} and
         also about $10^{51}$ erg in visible energy, such as in shock
\index{supernova!shock wave}
         waves in ordinary matter, the former stellar envelope
         and interstellar gas.  These
         neutrinos have an energy in the range of a few MeV to about 20
         MeV.  When stars are in stellar binary systems, they can also
         explode at low mass, but this process is believed to give only 10
         \% or less of all stellar explosions.  There appears to be a
connection to Gamma Ray Bursts (GRBs), but the physical details are far from
\index{$\gamma-$ray burst (GRB)}
clear at present; some suggest a highly anisotropic explosion, others an
explosion running along a pre-existing channel.  It is noteworthy that
above an original stellar mass of about 15 solar masses, stars also
         have a strong stellar wind, which for original masses above 25
         solar masses becomes so strong, that it can blow out most of the
         original stellar mass, even before the star explodes as a
         supernova.  The energy in this wind, integrated over the
         lifetime of the star, can attain the energy of the subsequent
         supernova, as seen in the shockwave of the explosion.

\subsection{Gamma Ray Bursts} \index{$\gamma-$ray burst (GRB)|(} 

Bursts of gamma ray emission \cite{1:Piran99b} come from the far
reaches of the universe, and are almost certainly the result of the
creation of a stellar mass black hole.  The duration of these bursts
\index{black hole}
ranges from a fraction of a second to usually a few seconds, and
sometimes hundreds of seconds.  Some such GRBs have afterglows in
other wavelengths like radio, optical and X-rays, with an optical
brightness which very rarely comes close to being detectable with
standard binoculars.  The emission peaks near 100 keV in observable
photon energy, and appears to have an underlying powerlaw character,
suggesting non-thermal emission processes. See the contribution by
E.~Waxman in this volume for a detailed discussion.
\index{$\gamma-$ray burst (GRB)|)}

\subsection{Active Galactic Nuclei}  

When massive black holes accrete, then their
\index{black hole!supermassive}
         immediate environment, usually thought to be an accretion  disk
\index{accretion}
         and  a powerful relativistic jet (i.e. where the material
is ejected with a speed very close to the speed of light) emits a luminosity
         often far in excess of the emission of all stars in the host
         galaxy put together.  There is the proposal of a ``unified
scheme", which contains the elements of a black hole, an accretion
disk, a jet and a torus of surrounding molecular material.  The mass
range of these black holes appears to extend to $3\times10^9$ solar
masses.  As an example such black holes of a mass near
         $10^8$ solar masses have a size of order the diameter of the
         Earth orbit around the Sun, and their accretion can produce a
total emission of 1000 times
         that of all stars in our Galaxy.  When the emission of the jets
         gets very strong, and the jet very powerful, then the radio image
         of such a galaxy can extend to 300 kpc, or more, dissipating the
\index{radio galaxy!active galactic nucleus (AGN)}
\index{radio galaxy!jet, non-relativistic}
\index{radio galaxy!jet, relativistic}
\index{radio galaxy!lobe}
\index{radio galaxy!hot spot}
         jet in radio hot spots embedded in giant radio lobes, very
rarely to several Mpc.  The space 
         density of such radio galaxies, with powerful jets, hot spots
         and lobes, is low, less than 1/1000 of all galaxies, but on the
         radio sky they dominate due to their extreme emission.  The
activity is thought to be fed by inflow of gas and/or stars into the
black hole, maybe usually fuelled by galaxy-galaxy interaction
\cite{1:SM96}.  High energy particle interactions in active galactic nuclei
\index{radio galaxy!active galactic nucleus (AGN)}
and their surroundings may be detectable through the neutrino emission,
\index{radio galaxy!neutrino production}
even at cosmological distances \cite{1:LM2000}. See also the contribution
by G.~Pelletier in this volume.

\subsection{Topological Defects and Supermassive Particles}  
\index{topological defect (TD)|(}
\index{supermassive (X) particle|(}
\index{topological defect (TD)!neutrino production}
Particle accelerator experiments and the mathematical structure of the
Standard Model of the weak, electromagnetic and strong interactions
suggest that these forces should be unified at energies of about
$2\times10^{16}\,$GeV ($1\,$GeV$=10^9\,$eV)~\cite{1:gut}, 4-5 orders
\index{grand unified field theory (GUT)}
of magnitude above the highest energies observed in cosmic rays. The
relevant ``Grand Unified Theories'' (GUTs) predict the existence of X
particles with mass $m_X$ around the GUT scale of
$\simeq2\times10^{16}\,$GeV$/c^2$.\index{grand unification (GUT) scale}
If their lifetime is comparable or larger than the age of the Universe,
they would be dark matter candidates\index{supermassive (X) particle}
and their decays could contribute to cosmic ray fluxes at
\index{supermassive (X) particle!as source of ultra-high energy cosmic rays}
\index{decay!supermassive (X) particle}
\index{cosmic ray!flux}
the highest energies today, with an anisotropy pattern that reflects
\index{arrival direction distribution!anisotropy}
the expected dark matter\index{dark matter}
distribution~\cite{1:evans}. However, in many
GUTs supermassive particles are expected to have lifetimes not much
longer than their inverse mass, $\sim6.6\times10^{-41}(10^{16}\,{\rm
GeV}/m_Xc^2)\,{\rm sec}$, and thus have to be produced continuously if
their decays are to give rise to ultra-high energy cosmic rays.  This
\index{topological defect (TD)!as source of ultra-high energy cosmic rays}
can only occur by emission from topological defects which are relics
of cosmological phase transitions that could have occurred in the
early Universe at temperatures close to the GUT scale.
\index{grand unification (GUT) scale}Phase
transitions in general are associated with a breakdown of a group of
symmetries down to a subgroup which is indicated by an order parameter
taking on a non-vanishing value. Topological defects occur between
regions that are causally disconnected, such that the orientation of
the order parameter cannot be communicated between these regions and
thus will adopt different values. Examples are cosmic
strings~\footnote{Strings correspond to the breakdown of rotational
symmetry $U(1)$ around a certain direction; a laboratory example are
vortices in superfluid helium.}, magnetic monopoles~\footnote{Magnetic
monopoles correspond to the breakdown of arbitrary 3-dimensional
rotations $SO(3)$ to rotations $U(1)$ around a specific direction.},
\index{topological defect (TD)!monopole}
\index{topological defect (TD)!cosmic string}
\index{topological defect (TD)!domain wall}
and domain walls~\footnote{Domain walls correspond to the breakdown of
a discrete symmetry where the order parameter is only allowed to take
several discrete values; a laboratory example are the Bloch walls
separating regions of different magnetization along the principal axis
of a ferromagnet.}. The Kibble mechanism states~\cite{1:kibble} that
about one defect forms per maximal volume over which the order
parameter can be communicated by physical processes.  The defects are
topologically stable, but in the case of GUTs time dependent motion
can lead to the emission of GUT scale\index{grand unification (GUT) scale}
X particles.

One of the prime cosmological motivations to postulate inflation, a
\index{inflation, cosmic}
phase of exponential expansion in the early Universe~\cite{1:kt}, was
to dilute excessive production of ``dangerous relics'' such as
topological defects and superheavy stable particles. However, right
after inflation, when the Universe reheats, phase transitions can
occur and such relics can be produced in cosmologically interesting
abundances where they contribute to the dark matter,\index{dark matter}
and with a mass scale roughly given by the inflationary
scale.  The mass scale is fixed by the CMB anisotropies to
\index{cosmic microwave background (CMB)}
$\sim10^{13}\,$GeV$/c^2$~\cite{1:kuz-tak}, and it is not far above the
highest energies observed in cosmic rays, thus motivating a connection
between these primordial relics and ultra-high energy cosmic rays
which in turn may provide a probe of the early Universe.

Within GUTs the X particles typically decay into jets of particles
\index{grand unified field theory (GUT)}
\index{decay!supermassive (X) particle}
whose spectra can be estimated within the Standard Model. Very
roughly, one expects a few percent nucleons and the rest in neutrinos and
\index{nucleon!primary}
\index{neutrino!primary}
\index{$\gamma-$ray!primary}
\index{supermassive (X) particle!as source of ultra-high energy cosmic rays}
\index{supermassive (X) particle!neutrino production}
\index{supermassive (X) particle!$\gamma-$ray production}
photons~\cite{1:bere-kachel}; these neutrinos and photons then cascade
\index{cascade!neutrino}
in the big bang relic
\index{neutrino!relic background (RNB)}
neutrinos and photons, and so produce a universal photon and neutrino
background (see the contribution by G.~Sigl in this volume). It is
not finally settled at which level we need to observe
a background to confirm or refute this expected background.  The
resulting hadron spectrum can be a fair bit flatter than any background
resulting from cosmic accelerators such as radio galaxies.  Therefore any
\index{radio galaxy}
background from the decay of topological defects or other relics should
produce observable signatures in neutrinos, photons and hadrons with
characteristic properties. For more details on the top-down
\index{top-down scenario}
scenario see the contribution by P.~Bhattacharjee and G.~Sigl in this volume.
\index{topological defect (TD)|)}
\index{supermassive (X) particle|)}

\subsection{Magnetic Monopoles}  \index{topological defect (TD)!monopole|(}

The physics of electric and magnetic fields
         contains electric charges but no magnetic charges.  In the
         context of particle physics it is likely that monopoles, basic
         magnetically charged particles, also exist.  Such monopoles are a
         special kind of topological defects.  The basic property of
         monopoles can be described as follows:  a)  Just as electrically
         charged particles shortcircuit electric fields, monopoles
         shortcircuit magnetic fields.  The observation of very large
         scale and permeating magnetic fields in the cosmos shows that the
\index{magnetic field!cosmic}
         universal flux of monopoles must be very low; the implied upper
limit from this argument is called the {\it Parker limit}.  b)  Monopoles
are accelerated in magnetic fields, just as electrically charged
         particles are accelerated in electric fields.  In cosmic magnetic
         fields, the energies which can be attained are of 
         $10^{21}$ eV, or even more.  Any relation to the observed high
         energy cosmic rays is uncertain at present~\cite{1:wkwb}.

\index{topological defect (TD)!monopole|)}

\subsection{Primordial Black Holes and Z-bursts}

In the early universe it is possible, that very small black holes were
also formed.  At sufficiently small mass, they can decay, and produce a
characteristic spectrum of particles rather similar to topological
defects~\cite{1:barrau}.
\index{black hole!primordial}
\index{decay!topological defect}

Another way to obtain very energetic hadrons is to start with
a neutrino at very high energy and at distances possibly much larger
\index{neutrino!primary}
than the energy loss lengths $\sim50\,$Mpc for photons, nucleons, and
\index{nucleon!energy loss}
\index{nucleon!attenuation length}
\index{cosmic ray!energy loss}
nuclei and have it interact with the relic neutrino background, the
\index{neutrino!relic background (RNB)}
neutrino analogue of the CMB~\cite{1:Zburst},
\index{cosmic microwave background (CMB)}
within $\sim50\,$Mpc. Such neutrino-neutrino interactions
\index{neutrino!interaction}
produce a Z boson, a carrier of the electroweak interactions,
which immediately decays into hadrons and other
particles, thus producing a proton possibly quite near to us in the
Universe. For more details on this ``Z-burst'' mechanism
\index{Z-burst scenario}
see the contributions by G.~Sigl and by S.~Yoshida on neutrino cascades
\index{cascade!neutrino}
in this volume.

\section{Energies, Spectra, and Composition}

The solar wind prevents low energy charged particles to come into the inner
\index{solar!wind}
solar system, due to interaction with the magnetic field in the solar
\index{magnetic field!solar wind}
wind, a steady stream of gas going out from the Sun into all directions,
originally discovered in 1950 from the effect on cometary tails:  they
all point outwards, at all latitudes of the Sun, and independent on
whether the comet actually comes into the inner solar system, or goes
outwards, in which case the tail actually precedes the head of the
comet.  This prevents us from knowing anything about the energies
lower than about 300 MeV of interstellar energetic particles.  From about
10 GeV per charge unit $Z$ of the particle, the effect of the solar wind
becomes negligible.  Since cosmic ray particles are mostly fully
ionized nuclei (i.e. with the exception of electrons and positrons), this
is a strong effect.

Our Galaxy has a magnetic field of about $6 \times 10^{-6}$ Gauss in the
\index{magnetic field!galactic}
solar neighbourhood; the energy of such a field corresponds approximately
to 1 eV per cm$^3$, just like the other components of the interstellar
medium.  In such a magnetic field charged energetic particles gyrate, with
a radius of gyration, called the Larmor radius, which is proportional to
\index{Larmor!radius}
the momentum of the particle perpendicular to the magnetic field
direction.  For highly relativistic particles this entails, that around
$3 \times 10^{18}$ eV protons - or other nuclei of the same energy  to
\index{deflection, magnetic}
charge ratio - no longer gyrate in the disk of the Galaxy, i.e. their
radius of gyration is larger than the thickness of the disk.   So they
cannot possibly originate in the Galaxy, they must come from outside; and
indeed, at that energy there is evidence for a change both in chemical
\index{cosmic ray!composition}
composition, and in the slope of the spectrum.
\index{cosmic ray!spectrum}

The energies of these cosmic ray particles, that we observe, range from a
\index{cosmic ray!spectrum}
few hundred MeV to $\simeq300\,$EeV. The integral flux ranges from
\index{cosmic ray!flux}
about $10^{-5}$ per cm$^2$, per s, per steradian, at 1 TeV per nucleus
for Hydrogen, or
protons, to 1 particle per steradian per km$^2$ and per century around
$10^{20}$ eV, a decrease by a factor of $3 \times 10^{19}$ in  integral
flux, and a corresponding decrease by a factor of $3 \times 10^{27}$ in
differential flux, i.e. per energy interval (see also Fig.~\ref{1:F1}).
Electrons have only been measured to a few TeV.

As already discussed in Sect.~\ref{1:s1},
the total particle spectrum is about $E^{-2.7}$ below the knee,
\index{cosmic ray!spectrum}
\index{knee, cosmic ray spectrum}
and about $E^{-3.1}$ above the knee, at 5 PeV, and flattens again to
about $E^{-2.7}$ beyond the ankle, at about 3 EeV.  Electrons have a
\index{ankle, cosmic ray spectrum}
spectrum, which is similar to that of protons below about 10 GeV, and
steeper, near $E^{-3.3}$ above this energy.  The lower spectrum of
electrons is inferred from radio emission, while the steeper spectrum at
the higher energies is measured directly.

The chemical composition is rather close to that of the interstellar
\index{cosmic ray!composition}
medium, with a few strong peculiarities relative to that of the
interstellar medium, see Sect.~\ref{1:s2} for a general discussion.
Concerning the energy dependence towards the
knee, and beyond, the fraction of heavy elements appears to continuously
\index{knee, cosmic ray spectrum}
\index{heavy nucleus!primary}
increase, with moderately to heavy elements almost certainly dominating
beyond the knee~\cite{1:kascade}, all the way to the ankle, where
\index{ankle, cosmic ray spectrum}
the composition seems to become light again~\cite{1:fe}.
This means, at that energy we observe a transition to what
appears to be mostly Hydrogen and Helium nuclei.  At much higher energies
we can only show consistency with a continuation of these properties, we
cannot prove unambiguously what the nature of these particles is.

The fraction of antiparticles is a few percent for positrons and
a few $10^{-4}$ for anti-protons. No other anti-nuclei have been
found~\cite{1:antimatter}.
\index{antimatter}

\section{Origin of Galactic Cosmic Rays}

\subsection{Injection}
\index{injection|(}

For the injection of cosmic rays the following reasons have been
suggested, and we will group the answers into three segments following
the very different paths of arguments.  

There is first the suggestion, that low mass stars with their coronal
activity provide the injection mechanism (mostly due to
M. Shapiro, \cite{1:Shapiro99}).  The
main argument for this reasoning is the observation that the selection
effects for the different elements among energetic particles are very
similar in the solar wind and in cosmic rays.  Since low mass stars are
\index{solar!wind}
often observed to be very active, their possible contribution is
expected to be substantial.  In fact, in a few other stars, these
selection effects have been checked \cite{1:Schmitt94,1:Drake95}.

The argument then proceeds as follows:

\begin{itemize}

\item{}  Low mass stars in their coronal activity accelerate selectively
certain elements to supra-thermal energies, and so inject them into the
interstellar medium.

\item{}  Normal supernova explosions then accelerate them, via shock
\index{supernova!shock wave}
waves running through the interstellar medium.

\end{itemize}

There is second the suggestion that the injection of cosmic rays starts
with ionized dust particles, and finishes by a break-up of the energetic
dust.  Many of the selection effects governing dust formation, and also
the sites of dust formation then rule the abundances of the final cosmic
ray particles.  

\begin{itemize}

\item{}  This model has been developed on the one hand by Luke
Drury and his collaborators~\cite{1:JPMeyer97}, and on the other by
the group of the late Reuven Ramaty and his collaborators~\cite{1:R97a}.

\item{}  One of the biggest successes of this theory is the rather good
explanation for the various abundances of the chemical elements just using
the known properties of dust, and the observed fact that dust is abundant
\index{cosmic ray!composition}
everywhere.

\item{}  A challenging aspect is the possibility to explain the
observational fact that the light elements such as Boron were already
abundant at early times in the Galaxy, when the general abundances of all
heavy elements were low; dust is formed early around the supernovae
\index{supernova} of
massive stars, such as supernova 1987a, as observations clearly indicate,
and so the general abundance of dust in the interstellar medium is of no
significance.  This aspect is one of the strengths of the approach by 
Ramaty. He elegantly solves the problem of the abundances of the light
elements in the young Galaxy.

\item{}  The isotopic ratios of certain elements clearly suggest that at
least some massive stars, such as Wolf Rayet stars, do contribute at some
level.  However, in this approach, they play a minor role.

\end{itemize}
 
There is a third, competing theory, which emphasizes the role played by
the very massive stars, and their winds.

\begin{itemize}

\item{}  Here the difference is noted, that massive stars come in
three well-understood varieties, i) those with a zero age main
sequence mass between 8 and 15 solar masses, which explode into the
interstellar medium, ii) those with a mass between 15 and about 25 solar
masses, which explode into their stellar wind, which is enriched mostly in
Helium, and finally those with a mass above about 25 solar masses, which
explode as blue supergiants, Wolf Rayet stars, for which the wind is
heavily enriched in Carbon and Oxygen.

\item{}  The interstellar turbulence spectrum is taken to be of Kolmogorov
\index{magnetic field!turbulence spectrum}
type~\cite{1:kolmogorov}, as indicated by an abundance of observations
and theoretical work \cite{1:Rickett90}.

\item{}  The injection happens from the stellar wind abundances,
explaining the general features of the abundances.  However, since some
elements are doubly ionized, their injection is enhanced, leading to
a selection effect well known from the active zones of the Sun and the
solar wind, and also seen in some active stars.  Therefore, this picture
\index{solar!wind}
also uses the analogy between the solar wind, and assumes that similar
selection effects play a role in the winds of massive stars.

\end{itemize}
\index{injection|)}

\subsection{Primary Acceleration}

It has been long surmised that supernova explosions provide the bulk
of the acceleration of cosmic rays in the Galaxy \cite{1:BZ34}.  The
\index{acceleration}
acceleration is thought to be a kind of ping-pong between the two
sides of the strong shock wave sent out by the explosion of the star.
\index{supernova!shock wave}
\index{acceleration!diffusive shock}
This ping pong is a repeated reflection via magnetic resonant
interaction between the gyromotion of the energetic charged particles,
and waves of the same wavelength as the Larmor motion in the magnetic
\index{Larmor!radius}
thermal gas.  Since the reflection is usually thought to be a gradual
diffusion in direction, the process is called diffusive shock
\index{diffusion!spatial}
acceleration, or after its discoverer
Fermi acceleration~\cite{1:Fermi}; see the contribution by
G.~Pelletier in this volume for a detailed discussion.

For a shock wave sent out directly into the interstellar gas this kind of
acceleration easily provides particle energies up to about 100 TeV. 
While the detailed injection mechanism is not quite clear, the very fact
that we observe the emission of particles at these energies in X-rays
provides a good case, and a rather direct argument for highly energetic
electrons.  Even though protons are by a factor of about 100 more abundant
at energies near 1 GeV than electrons, we cannot prove yet directly that
supernova shocks provide the acceleration; only the analogy with electrons
\index{supernova!shock wave}
can be demonstrated.
\index{acceleration}
\index{acceleration!maximum energy}

However, we observe what are probably Galactic cosmic rays up to energies
near the knee, and beyond to the ankle, i.e. 3 EeV.
\index{knee, cosmic ray spectrum}
\index{ankle, cosmic ray spectrum}

The energies can be provided by several possibilities, with the only
theory worked out to a quantitative level suggesting that those particles
also get accelerated in supernova shock waves, in those which run
\index{supernova!shock wave}
through the powerful stellar wind of the predecessor star.  In this first
possibility it can easily be shown, that energies up to 3 EeV per particle
are possible (mostly Iron then).  An alternate, second, possibility is
that a ping pong between various supernova shockwaves occurs, but in this
case seen from outside.  In either (or any other) such theory it is a
problem, that we observe a knee, i.e. a bend down of the spectrum at an
\index{knee, cosmic ray spectrum}
energy per charge ratio which appears to be fairly sharply defined.  In
the concept (the first possibility) that stellar explosions are at the
origin it entails that all such stars are closely similar in their
properties, including their magnetic field, at the time of explosion;
while this is certainly possible, we have too little information on the
magnetic field of pre-supernova stars to verify or falsify this.  In the
\index{magnetic field!stellar}
case of the other concept (the second possibility) it means that the
transport through the interstellar gas has a change in properties also at
\index{cosmic ray!transport}
a fairly sharply defined energy to charge ration, indicating a special
scale in the interstellar gas, for which there is no other evidence.

Galactic cosmic rays get injected from their sources with a certain
spectrum.  While they travel through the Galaxy, from the site of
injection to escape or to the observer, they have a certain chance to leak
\index{cosmic ray!escape}
out from the hot galactic magnetic disk of several kpc thickness.
This escape becomes easier with higher energy.  As a consequence their
spectrum steepens, comparing source and observed spectrum.  The radio
observations of other galaxies show consistency with the understanding
that the average spectrum of cosmic rays at least in the GeV to many GeV
energy range is always the same, in various locations in a Galaxy, and
also the same in different galaxies.  During this travel inside a galaxy
the cosmic rays interact with the interstellar gas, and in this
interaction produce gamma ray emission from pion decay, positrons, and
\index{$\gamma-$ray!secondary}
also neutrons, anti-protons, and neutrinos.  The future gamma ray emission
\index{neutrino!secondary}
observations will certainly provide very strong constraints on this
aspect of cosmic rays.
\index{cosmic ray!spectrum}

One kind of evidence where cosmic rays exactly come from, what kind of
stars and stellar explosions really dominate among their sources is
the isotopic ratios of various isotopes of Neon, Iron and other heavy
\index{heavy nucleus!primary}
elements; these isotope ratios suggest that at least one population is
indeed the very massive stars with strong stellar winds; however, whether
these stars provide most of the heavier elements, as one theory proposes,
is still quite an open question. 

There is some evidence now, that just near EeV energies there is one
component of galactic cosmic rays, which is spatially associated in
arrival direction with the two regions of highest activity in our Galaxy,
\index{arrival direction distribution}
at least as seen from Earth (by AGASA and SUGAR):  the Galactic Center
\index{SUGAR ground array}
\index{Akeno Giant Air Shower Array (AGASA)}
region as well as the Cygnus region show some weak
enhancement~\cite{1:clayetal}. Such a
directional association is only possible for neutral particles, and since
neutrons at that energy can just about travel from those regions to here,
before they decay (only free neutrons decay, neutrons bound into a
nucleus do not decay), a production of neutrons is conceivable as one
explanation of these data.  One major difficulty with this
interpretation is the lack of discernible high energy gamma ray emission
\index{$\gamma-$ray!secondary}
associated with the regions of presumed neutron emission; the CASA-MIA
experiment only provided stringent upper limits~\cite{1:borione}, which
appear on first sight to
rule out the possibility that related interactions might provide the
neutrons.  On the other hand, these two regions are clearly those two
parts of the Galaxy, where cosmic ray interactions are the strongest, as
evidenced by both lower energy gamma data as well as radio data.
\index{$\gamma-$ray!secondary}

\subsection{Beyond the Knee}
\index{knee, cosmic ray spectrum|(}

There are several ideas how to get particles accelerated to energies near
and beyond the knee, at about $5\times10^{15}$ eV.  The observations of air
showers suggest that the knee is a feature in constant energy per charge,
or rigidity, as surmised already by B. Peters~\cite{1:Peters}. 
\index{rigidity, magnetic}
The same may be true of the ``second knee", near $3\times10^{17}$ eV.

There are again several approaches conceivable, with only one
quantitative theory for this energy range:

\begin{itemize}

\item{}  Obviously, a new accelerator, such as pulsars, might take over;
\index{pulsar!as source of ultra-high energy cosmic rays}
however, then the steeper spectrum with a matching flux at the knee
\index{cosmic ray!flux}
energy is  a serious problem, and so this notion is normally discounted
today.
\index{cosmic ray!spectrum}

\item{}  In the context of the injection from energetic particles from
\index{injection}
low mass active stars, an additional unidentified process provides
further acceleration to those energies beyond the knee.
\index{acceleration}

\item{}  In the model using dust particles as primary injection
\index{injection}
mechanism there is no account of the cosmic ray spectrum beyond the
\index{cosmic ray!spectrum}
knee.  A development of the theory, using acceleration between the
expanding shells and shocks of different supernovae might solve this
\index{supernova!shock wave}
problem.
\index{acceleration}

\item{}  In the theory using the supernova shock racing through stellar
\index{supernova!shock wave}
winds, their shell, and the immediate surroundings, all particle
energies up to the ankle can be explained due to shock acceleration in
\index{acceleration}
\index{ankle, cosmic ray spectrum}
the wind, which is magnetized.  The knee is explained as due to a
\index{magnetic field!stellar}
\index{magnetic field!supernova remnant}
diminution of the acceleration efficiency when drift acceleration is
reduced due to the matching of the Larmor radius of the motion of the
\index{Larmor!radius}
particle, and the spatial constraints in a shocked shell, racing
through the stellar wind.

\end{itemize}

\index{knee, cosmic ray spectrum|)}
\subsection{Transport in the Galaxy}

Cosmic ray particles are diffusively transported through the Galaxy,
\index{diffusion!spatial}
\index{cosmic ray!transport}
interacting all the time with the matter, magnetic fields and photons. 
\index{magnetic field!galactic}
The various theories differ in which interaction site dominates.

\begin{itemize}

\item{} In the theory using dust particles the injection is with a spectrum of
\index{injection}
$E^{-2.1}$ approximately, and so an interstellar turbulence spectrum such that it
\index{magnetic field!turbulence spectrum}
would lead to a steepening in $E^{-0.6}$ is required, for which there is little
convincing observational nor theoretical evidence, except indirectly through using an
adopted model of a leaky box for cosmic ray transport.  Again, a further development
\index{leaky-box model}
\index{cosmic ray!transport}
of the theory might remedy this aspect.  Especially, re-acceleration in the
\index{re-acceleration, cosmic rays}
interstellar medium might help, as argued by Seo and Ptuskin~\cite{1:SeoPt94}.

\item{}  In the theory using stellar winds the cosmic ray interaction 
happens in the shells around the stellar winds \cite{1:TucsonCR,1:CRIX}, and
their immediate environments, explaining readily the energy dependence of
the ratio of the secondary elements from spallation and the primary
\index{secondary/primary ratio}
elements, with
$E^{-5/9}$.  This also explains the gamma ray spectrum, which is observed
\index{$\gamma-$ray!secondary}
to be best approximated by an interaction spectrum of $E^{-2.3}$.  And,
furthermore, this approach also explains the electron spectrum, observed
\index{cosmic ray!spectrum}
to be $E^{-3.3}$, and since it is dominated by losses, requires an
\index{cosmic ray!energy loss}
injection close to a spectrum of $E^{-2.3}$, as noted earlier.
\index{injection}

\end{itemize}

For an example for detailed modeling of cosmic ray progagation
and secondary production in the Galaxy see, e.g., Ref.~\cite{1:moskalenko}.

\subsection{Key Tests}

In all these theories, there are critical aspects which are not yet
developed, and will surely determine in the future, which of these
proposals, if anyone of them, does explain what Nature is doing.

\begin{itemize}

\item{}  In the picture using energetic particles from low mass active
stars a key test would be the isotopic abundances, comparing those in the
solar wind, and those in cosmic rays.
\index{solar!wind}

\item{}  In the theory using dust particle injection the expected gamma
\index{injection}
\index{$\gamma-$ray!secondary}
ray spectrum from cosmic ray interactions has not been worked out yet, and
may finally confirm this approach, or falsify it.  Also, the isotopic
abundances provide key tests.

\item{} What has yet to be done, and may well finally prove or falsify
the theory involving stellar winds is the very detailed accounting of all
the abundances of the chemical elements and their isotopic abundances.
\index{cosmic ray!composition}

\item{}  And, finally, once we observe the high energy gamma ray emission
\index{$\gamma-$ray!secondary}
spectrum, its spatial distribution, as well as the neutrino spectrum from
\index{neutrino!spectrum}
the inner part of our Galaxy, then we can expect to finalize our
physical understanding of where cosmic rays come from.

\end{itemize}

Observations such as \cite{1:CASAMIA} may provide key tests for
progress from the knee on up.
\index{knee, cosmic ray spectrum}

\section{The Cosmic Rays between 3 EeV and 50 EeV}

The cosmic rays between the ankle and the expected GZK-cutoff are readily
\index{Greisen-Zatsepin-Kuzmin effect (GZK)}
explained by many possible sources, almost all outside our galaxy.  
\index{ankle, cosmic ray spectrum}

Some, but not all of these proposals can also explain particles beyond
the GZK-cutoff, discussed in Sect.~\ref{1:s3} below.
\index{Greisen-Zatsepin-Kuzmin effect (GZK)}

Pulsars, especially those with very high magnetic fields, called
\index{pulsar!as source of ultra-high energy cosmic rays}
\index{magnetic field!pulsar}
\index{pulsar!as source of ultra-high energy cosmic rays}
magnetars, can 
%certainly 
possibly accelerate charged particles to energies of $10^{21}$ eV (see
contribution by B.~Rudak in this volume). There are several problems
with such a notion, one being the adiabatic losses on the way from
\index{cosmic ray!energy loss}
\index{adiabatic energy loss}
close to the pulsar out to the interstellar gas, and another one the
sky distribution, which should be 
%very 
\index{arrival direction distribution!anisotropy}
anisotropic given the distribution and strength of Galactic magnetic
\index{magnetic field!galactic}
fields.  On the other hand if this concept could be proven, it would
certainly provide a very easy explanation, why there are particles
beyond the GZK-cutoff: for Galactic particles the interaction with the
\index{Greisen-Zatsepin-Kuzmin effect (GZK)}
CMB is totally irrelevant, and no GZK-cutoff is expected.
\index{cosmic microwave background (CMB)}

Another proposal is GRBs, and is discussed in detail in the
\index{$\gamma-$ray burst (GRB)}
contribution by E.~Waxman in this volume. However since ultimately we
do not yet know what constitutes a GRB, their contribution cannot be
settled with full certainty.

Shock waves running through a magnetized and ionized gas accelerate
charged particles, as we know from in situ observations in the solar
\index{solar!wind}
wind already; and this forms the basis of almost all theories to
account for Galactic Cosmic Rays.  The largest shock waves in the
universe have scales of many tens of Mpc, and have shock velocities of
\index{shock!large scale structure accretion}
around 1000 km/s.  These shock waves arise in the cosmological large
scale structure formation, seen as a soap-bubble like distribution of
galaxies in the universe.  The accretion flow to enhance the matter
\index{accretion}
density in the resulting sheets, filaments and clusters is still
\index{galaxy!cluster}
continuing, and causes shock waves to exist all around us.  In the
shock waves, which also have been shown to form around growing
clusters of galaxies, particles can be accelerated, and can attain
fairly high energies.  However, the maximum energies can barely reach
\index{acceleration!maximum energy}
the energy of the GZK-cutoff, and so a strong contribution to the
\index{Greisen-Zatsepin-Kuzmin effect (GZK)}
overall flux is unlikely~\cite{1:kangetal}.
\index{cosmic ray!flux}

The most conventional 
%and easiest 
explanation is radio galaxies, which provide with their hot spots an
\index{radio galaxy!as source of ultra-high energy cosmic rays}
\index{radio galaxy!hot spot}
obvious acceleration site: These hot spots are giant shock waves,
\index{acceleration}
often of a size exceeding that of our entire Galaxy.  The shock speeds
\index{shock!velocity}
may approach several percent, maybe even several tens of percent of
the speed of light, if sporadic.  
%Therefore, in this interpretation,
%these radio galaxy hot spots provide a very straightforward
%acceleration site.  
Integrating over all known radio galaxies readily explains flux and
\index{cosmic ray!flux}
spectrum, as well as chemical composition of the cosmic rays in this
\index{cosmic ray!composition}
energy range~\cite{1:GS63,1:Hillas84,1:BS87}.  In this proposal it is
the greatest challenge to identify the single radio galaxy dominating
the highest energy; for this M87 has been proposed already some time
ago (see also the contribution by P.~Biermann et al. in this volume).

\section{Particles beyond the GZK-cutoff}\label{1:s3}
\index{Greisen-Zatsepin-Kuzmin effect (GZK)}

For these energies there is no argument, whether these particles are
really protons, as an extrapolation from lower energies might suggest. 
However, everything we know is quite consistent with such an
assumption~\cite{1:NW2000}.

Apart from the more ``conservative'' astrophysical mechanisms
involving \linebreak
``bottom-up'' acceleration, there are many exciting approaches
\index{acceleration}
\index{bottom-up scenario}
to account for these particles:

\begin{itemize}

\item{}  Decay of topological defects (TDs), or other relics from the big
bang, the so-called ``top-down'' scenario. This theory can account
\index{top-down scenario}
readily for the apparent upturn in the spectrum beyond the GZK cutoff,
\index{Greisen-Zatsepin-Kuzmin effect (GZK)}
and explains those events with a mixture of nucleons and
\index{nucleon!primary}
$\gamma-$rays. These models predict significant diffuse $\gamma-$ray
\index{$\gamma-$ray!primary}
\index{$\gamma-$ray!secondary}
fluxes in the 100 MeV-GeV region and thus are strongly constrained by
the observed fluxes in this energy range. There are many variants of
top-down models~\cite{1:bs}, some of them with a quite predictive
power.

\item{}  Decay of primordial black holes.  The final
\index{black hole!primordial}
particle distribution is rather similar to that expected from the
decay of TDs~\cite{1:barrau}.
\index{decay!topological defect}

\item{}  Violation of the Lorentz invariance~\cite{1:linvar}:
\index{Lorentz invariance!violation}
At some very high energy, where the four basic
forces of Nature combine, Lorentz Invariance may no longer hold,
and a ripple effect of this is anticipated at lower energies.  One
possible result would be that protons might survive much longer in
the bath of the CMB. In fact, observations of photons
\index{cosmic microwave background (CMB)}
of energies up to $\simeq20\,$TeV from Markarian 501, where absorption
in the infrared background is expected to be strong, was considered
\index{infrared (IR) background}
as a possible signature of violation of Lorentz invariance~\cite{1:ph,1:acp}.
Furthermore, photons at different energies would have
divergent travel times, conceivably measurable with GRBs~\cite{1:acp}.
\index{$\gamma-$ray burst (GRB)}
\index{Lorentz invariance!test}

\end{itemize}

\section{Outlook}

The next few years promise to give great advances to our physical
understanding of both the macro and the microcosmos. On the one
side, this is due to our increased theoretical understanding on
how to combine accelerator data and cosmic ray and astrophysical
data to arrive at strong constraints, for example, on new physics.
On the other hand, it is due to an expected enormous increase
of data from new experiments, especially on the cosmic ray and
astrophysics side. Ground arrays, Balloons, Space Station experiments
will proliferate within the next few years and hold great promise for us.
On a somewhat longer time scale, powerful new particle accelerators
such as the LHC will directly test new physics in the TeV region,
an energy range which is also, somewhat more indirectly, probed
by cosmic ray, $\gamma-$ray and neutrino experiments.

\subsection*{Acknowledgement}

Work on these topics with PLB has been supported by grants from NATO, the
BMBF-DESY, the EU, the DFG and other sources.  PLB has to thank many
people, especially all those with whom he has recently published and
worked in this area, or is doing it at present, especially Eun-Joo Ahn,
Alina Donea, Torsten Ensslin, Heino Falcke, Glennys Farrar, Martin Harwit,
Hyesung Kang, Tom Kephardt, Phil Kronberg, Norbert Langer, Sera Markoff,
Gustavo Medina-Tanco, Ray Protheroe, Giovanna Pugliese, Wolfgang
Rhode, Dongsu Ryu, Eun-Suk Seo, Ramin Sina, Todor Stanev, Samvel
Ter-Antonyan, Amri Wandel, Yiping Wang, Tom Weiler, Stefan
Westerhoff, and of course, the present coauthor. GS is especially
grateful to the late David Schramm for his continuous encouragment
to work on the exciting subject of ultrahigh energy cosmic rays.
GS also wishes to thank his recent collaborators on this subject,
Luis Anchordoqui, Phijushpani Bhattacharjee, Paolo Coppi, John Ellis, 
Christopher Hill, Claudia Isola, Karsten Jedamzik, Sangjin Lee,
Martin Lemoine, Angela Olinto, Gustavo Romero, Diego Torres, Craig Tyler,
Shigeru Yoshida, and, of course, the current coauthor.


\begin{thebibliography}{999}
\addcontentsline{toc}{section}{References}

\bibitem{1:CRA} V.~F.~Hess: Phys.~Z. {\bf 13}, 1084 (1912)

\bibitem{1:CRB} W.~Kohlh\"orster: Phys.~Z. {\bf 14}, 1153 (1913)

\bibitem{1:fe} D.~J.~Bird, et al.: Phys.~Rev.~Lett. {\bf 71},
3401 (1993)\\
D.~J.~Bird, et al.: Astrophys.~J. {\bf 424}, 491 (1994)\\
D.~J.~Bird, et al.: Astrophys.~J. {\bf 441}, 144 (1995)

\bibitem{1:haverah} See, e.g., M.~A.~Lawrence, R.~J.~O.~Reid,
A.~A.~Watson: J.~Phys.~G Nucl.~Part.~Phys. {\bf 17}, 733 (1991), and
references therein\\
see also {\sf http://ast.leeds.ac.uk/haverah/hav-home.html}

\bibitem{1:agasa} N.~Hayashida, et al.: Phys.~Rev.~Lett. {\bf 73}, 3491
(1994)\\
S.~Yoshida, et al.: Astropart.~Phys. {\bf 3}, 105 (1995)\\
M.~Takeda, et al.: Phys.~Rev.~Lett. {\bf 81}, 1163 (1998)\\
see also {\sf http://icrsun.icrr.u-tokyo.ac.jp/as/project/agasa.html}

\bibitem{1:icrr90} Proc. International Symposium on {\it Astrophysical 
Aspects of the Most Energetic Cosmic Rays}, eds. M.~Nagano and
F.~Takahara (World Scientific, Singapore, 1991)

\bibitem{1:icrr96} Proc. of
International Symposium on {\it Extremely High Energy Cosmic Rays:
Astrophysics and Future Observatories}, ed. M.~Nagano (Institute
for Cosmic Ray Research, Tokyo, 1996)

\bibitem{1:yakutsk} N.~N.~Efimov, et al.: Ref.~\cite{1:icrr90}, p.~20\\
B.~N.~Afnasiev: in Ref.~\cite{1:icrr96}, p.~32. 

\bibitem{1:volcano} J.~Linsley: Phys.~Rev.~Lett. {\bf 10}, 146 (1963)\\
J.~Linsley: Proc.~{\it 8th International Cosmic Ray Conference} {\bf 4},
295 (1963)

\bibitem{1:sugar} R.~G.~Brownlee, et al.: Can.~J.~Phys. {\bf 46}, S259
(1968)\\
M.~M.~Winn, et al.: J.~Phys.~G {\bf 12}, 653 (1986)\\
see also {\sf http://www.physics.usyd.edu.au/hienergy/sugar.html}

\bibitem{1:Cocconi56} G.~Cocconi: Nuovo Cimento, 10th ser.
{\bf 3}, 1433 (1956)

\bibitem{1:greisen} K.~Greisen: Phys.~Rev.~Lett. {\bf 16},
748 (1966)

\bibitem{1:zat-kuz} G.~T.~Zatsepin, V.~A.~Kuzmin,
Pis'ma~Zh.~Eksp.~Teor.~Fiz. {\bf 4}, 114 (1966) [JETP. Lett. {\bf 4},
78 (1966)]

\bibitem{1:AMS} AMS collaboration: Phys.~Lett. {\bf B490}, 27 (2000)\\
AMS collaboration: Phys.~Lett. {\bf B494}, 193 (2000)

\bibitem{1:iss} See {\sf
http://spaceflight.nasa.gov/station/science/space/index.html}

\bibitem{1:icrc24} Proc.~{\it 24th International Cosmic Ray Conference}
(Istituto Nazionale Fisica Nucleare, Rome, Italy, 1995)

\bibitem{1:hires} S.~C.~Corbat\'{o}, et al.: Nucl.~Phys.~B
(Proc. Suppl.) {\bf 28B}, 36 (1992)\\
D.~J.~Bird, et al., in Ref.~\cite{1:icrc24},
Vol.~{\bf 2}, 504; Vol.~{\bf 1}, 750\\
M.~Al-Seady, et al., in Ref.~\cite{1:icrr96}, p.~191\\
see also {\sf http://hires.physics.utah.edu/}

\bibitem{1:auger} J.~W.~Cronin: Nucl.~Phys.~B (Proc.~Suppl.) {\bf 28B},
213 (1992)\\
The Pierre Auger Observatory Design Report (2nd edition), March 1997\\
see also {\sf http://www.auger.org/} and
{\sf http://www-lpnhep.in2p3.fr/auger/welcome.html}

\bibitem{1:data} S.~Yoshida, H.~Dai: J.~Phys.~G {\bf 24}, 905
(1998)\\
X.~Bertou, M.~Boratav, A.~Letessier-Selvon:
Int.~J.~Mod.~Phys. {\bf A15}, 2181 (2000), and references therein.

\bibitem{1:NW2000}  M.~Nagano, A.~A.~Watson: Rev.~Mod.~Phys. {\bf 72}, 689
(2000)
%  Observations and implications of the ultra-high energy cosmic
%  rays

\bibitem{1:Fermi} E.~Fermi: Phys.~Rev. 2nd ser., {\bf 75}, no. 8,
1169 (1949)\\
E.~Fermi: Astrophys.~J. {\bf 119}, 1 (1954)

\bibitem{1:G53a}  V.~L.~Ginzburg: Usp.~Fiz.~Nauk {\bf 51}, 343 (1953)\\
See also {\it Nuovo Cimento}  {\bf 3}, 38 (1956)
%the origin of cosmic rays and radio astronomy

\bibitem{1:G53b} V.~L.~Ginzburg: Dokl.~Akad.~Nauk~SSSR {\bf 92}, 1133 (1953)
(NSF-Transl. 230)

\bibitem{1:GS63}  V.~L.~Ginzburg, S.~I.~Syrovatskii:
{\it The origin of cosmic rays} (Pergamon Press, Oxford 1964),
Russian edition (1963)

\bibitem{1:Gaisser90}  T.~K.~Gaisser: {\it Cosmic Rays and Particle
Physics} (Cambridge Univ. Press 1990)

\bibitem{1:G93} V.~L.~Ginzburg: Phys.~Usp. {\bf 36}, 587 (1993)
%The origin of cosmic rays - forty years later

\bibitem{1:G96}  V.~L.~Ginzburg: Uspekhi Fizicheskikh Nauk
{\bf 166}, 169 (1996)
%  Cosmic ray astrophysics (history and general review)

\bibitem{1:Venyabook} V.~S.~Berezinskii, et al.: {\it Astrophysics of
Cosmic Rays} (North-Holland, Amsterdam 1990), especially chapter IV.

\bibitem{1:Hires2000} T.~Abu-Zayyad, et al.: Astrophys.~J. {\bf 557}, 686
(2001)
%    Title: Measurement of the Cosmic Ray Energy Spectrum and Composition
%  from 10^{17} to 10^{18.3} eV 

\bibitem{1:Peters} B.~Peters: Nuovo Cimento Suppl. {\bf XIV}, 436 (1959)\\
B.~Peters: Nuovo Cimento {\bf XXII}, 800 (1961)

\bibitem{1:fe-aniso} D.~J.~Bird, et al.: Astrophys.~J. {\bf 511}, 739 (1999)

\bibitem{1:agasa-aniso} N.~Hayashida, et al.: Astropart.~Phys. {\bf 10},
303 (1999)

\bibitem{1:clayetal} J.~A.~Bellido, R.~W.~Clay, B.~R.~Dawson,
M.~Johnston-Hollitt: Astropart.~Phys. {\bf 15}, 167 (2001)

\bibitem{1:clustering} N.~Hayashida, et al.: Phys.~Rev.~Lett. {\bf 77}, 1000
(1996)\\
M.~Takeda, et al.: Astrophys.~J. {\bf 522}, 225 (1999); and astro-ph/0008102

\bibitem{1:BE87} R.~D.~Blandford, R.~D.~Eichler: Phys.~Rep. {\bf 154},
1 (1987)
%CR review

\bibitem{1:Drury83} L.~O'C~Drury: Rep.~Prog.~Phys. {\bf 46}, 973 (1983)
%  review on cosmic ray acceleration

\bibitem{1:Jokipii87}  J.~R.~Jokipii: Astrophys.~J. {\bf 313}, 842 (1987)
%  title = "Rate of energy gain and maximum energy in diffusive shock
%  acceleration"

\bibitem{1:BedO98} J.~Bednarz, M.~Ostrowski: Phys.~Rev.~Lett.
{\bf 80}, 3911 (1998)
%  Energy spectra of cosmic rays accelerated at
%  ultra-relativistic shock waves

\bibitem{1:GM77} M.~Garcia-Munoz, G.~M.~Mason, J.~A.~Simpson: Astrophys.~J.
{\bf 217}, 859 (1977)
%``The age of galactic cosmic rays derived from the abundance of
%$^{10}$Be"

\bibitem{1:GM87} M.~Garcia-Munoz, J.~A.~Simpson,
T.~G.~Guzik, J.~P.~Wefel, S.~H.~Margolis: Astrophys.~J.~Suppl.
{\bf 64}, 269 (1987)
%``Cosmic-ray propagation in the Galaxy and in the
% heliosphere:  The path-length distribution at low energy",

\bibitem{1:Tsao2001} C.~H.~Tsao, R.~Silberberg, A.~F.~Barghouty:
Astrophys.~J. {\bf 549}, 320 (2001)
%    title = "Cosmic-Ray Sources and Source Composition",

\bibitem{1:rhic} See {\sf http://www.bnl.gov/rhic/}

\bibitem{1:risks} See, e.g.:\\
R.~L.~Jaffe, W.~Busza, J.~Sandweiss, F.~Wilczek:
Rev.~Mod.~Phys. {\bf 72}, 1125 (2000)\\
A.~Dar, A.~De~R\'ujula, U.~Heinz: Phys.~Lett. {\bf B470}, 142 (1999)\\
A.~Kent, hep-ph/0009130

\bibitem{1:Reeves} H.~Reeves: Ann.~Rev.~Astron.~Astroph. {\bf 12}, 437
(1974)\\
H.~Reeves: Rev. Mod. Phys. {\bf 66}, 193 (1994)

\bibitem{1:LB-CR} B.~Wiebel-Sooth, P.~L.~Biermann:
Landolt-B{\"o}rnstein, vol. {\bf VI/3c}, Springer Verlag 1999, pp. 37 - 90
%  general review on cosmic rays physics, including long lists of 
%  experiments, a section on UHECR experiments is by A.A. Watson

\bibitem{1:R97a} R.~Ramaty, B.~Kozlovsky, R.~E.~Lingenfelter,
H.~Reeves: Astrophys.~J. {\bf 488}, 730 (1997)
% Light elements and cosmic rays in the early galaxy

\bibitem{1:euso} See {\sf http://www.ifcai.pa.cnr.it/Ifcai/euso.html}

\bibitem{1:icrc25} Proc.~25th {\it International Cosmic Ray Conference},
eds.: M.~S.~Potgieter, et al. (Durban, 1997)

\bibitem{1:owl} J.~F.~Ormes, et al.: in Ref.~\cite{1:icrc25},
Vol.~{\bf 5}, 273\\
Y.~Takahashi, et al.: in Ref.~\cite{1:icrr96}, p.~310\\
see also {\sf http://lheawww.gsfc.nasa.gov/docs/gamcosray/hecr/OWL/}

\bibitem{1:UHECR} J.~P.~Rachen, P.~L.~Biermann: Astron.~Astrophys.
{\bf 272}, 161 (1993)\\
J.~P.~Rachen, T.~Stanev, P.~L.~Biermann, Astron.~Astrophys.
{\bf 273}, 377 (1993)
%
\bibitem{1:GMT99b} G.~A.~Medina~Tanco: Astrophys.~J.~Lett. {\bf  510},
L91 (1999)
%  title = "The Energy Spectrum Observed by the AGASA Experiment and the
%  Spatial Distribution of the Sources of Ultra-High-Energy Cosmic Rays",

\bibitem{1:BBO2001} M.~Blanton, P.~Blasi, A.~V.~Olinto: Astropart.~Phys.
{\bf 15}, 275 (2001)

\bibitem{1:KR95}  J.~Kormendy, D.~Richstone:
Ann.~Rev.~Astron.~Astrophys. {\bf 33}, 581 (1995)
% search for central bh in galaxies
%

\bibitem{1:Faber97}  S.~M.~Faber, et al.: Astron.~J. {\bf 114}, 1771 (1997)
%  central black holes in galaxies, HST data

\bibitem{1:Magorrian98}  J.~Magorrian,, et al.:
Astron.~J. {\bf 115}, 2285 (1998)
% demography of central dark objects in galaxies

\bibitem{1:Gebhardt2000} K.~Gebhardt, et al.: Astrophys.~J.~Lett.
{\bf 539}, L13 (2000)
%  title = "A Relationship between Nuclear Black Hole Mass and Galaxy
%  Velocity Dispersion",

\bibitem{1:LSh78}  A.~P.~Lightman, S.~L.~Shapiro:
Rev.~Mod.~Phys. {\bf 50}, 437 (1978)
% evolution of globular clusters and central star clusters around bh

\bibitem{1:MF2001}  F.~Melia, H.~Falcke: Ann.~Rev.~Astron.~Astrophys.
(in press) (2001)
% review on the GC black hole

\bibitem{1:boomerang} C.~B.~Netterfield, et al.: astro-ph/0104460

\bibitem{1:maxima} A.~T.~Lee, et al.: astro-ph/0104459

\bibitem{1:dasi} N.~W.~Halverson, et al.: astro-ph/0104488, 0104489,
0104490

\bibitem{1:typeIa} S.~Perlmutter, et al.: Astrophys.~J. {\bf 517},
565 (1999)\\
A.~G.~Riess, et al.: Astron.~J. {\bf 116}, 1009 (1998)

\bibitem{1:combined} see also N.~A.~Bahcall, J.~P.~Ostriker, S.~Perlmutter,
P.~J.~Steinhardt: Science {\bf 284}, 1481 (1999)\\
R.~Stompor, et al.: astro-ph/0105062

\bibitem{1:Ostriker93} J.~P.~Ostriker: Ann.~Rev.~Astron.~Astrophys.
{\bf  31}, 689 (1993)
%  Title:  Astronomical tests of the cold dark matter scenario

\bibitem{1:Mateo98} M.~Mateo: Ann.~Rev.~Astron.~Astrophys. {\bf 36},
435 (1998)
%  Title:  Dwarf Galaxies of the Local Group

\bibitem{1:ROSAT97} S.~L.~Snowden, et al.: Astrophys.~J. {\bf 485}, 125 (1997)
%  ROSAT survey diffuse X-ray background, radial extent 5.6 kpc,
%  vertical scale height 1.9 kpc,
%  in plane electron density 0.0035 cm$^{-3}$, temperature $ 4 \, 106$ K.

\bibitem{1:Kaneda97}  H.~Kaneda, K.~Makishima, S.~Yamauchi, K.~Koyama,
K.~Matsuzaki, N.~Y.~Yamasaki: Astrophys.~J.  {\bf 491}, 638 (1997)
%  Title:  Complex Spectra of the Galactic Ridge X-Rays Observed with ASCA
%  ...very high T in ridge
%..abstract.." spectra are fairly well fitted by a double-temperature
%  nonequilibrium ionization plasma model with temperatures of kT ~ 0.8
%  keV and kT ~ 7 keV. .."

\bibitem{1:Valinia98} A.~Valinia, F.~E.~Marshall: Astrophys.~J.  {\bf 505}, 134
(1998)
%  Title:  RXTE Measurement of the Diffuse X-Ray Emission from the
%  Galactic Ridge: Implications for the Energetics of the Interstellar
%  Medium

\bibitem{1:Pietz98} J.~Pietz, J.~Kerp, P.~M.~W.~Kalberla, W.~B.~Burton,
D.~Hartmann, U.~Mebold: Astron.~Astrophys. {\bf 332}, 55 (1998)
%  Title:  The Galactic X-ray halo
%

\bibitem{1:Kronberg94} P.~P.~Kronberg: Rep.~Prog.~Phys., {\bf 57},
325 (1994)
%  Title:  Extragalactic magnetic fields

\bibitem{1:Beck96} R.~Beck, A.~Brandenburg, D.~Moss, A.~Shukurov, D.~Sokoloff:
Ann.~Rev.~Astron.~Astrophys. {\bf 34}, 155 (1996)
%  Title:  Galactic Magnetism:  Recent Developments and
%  perspectives

\bibitem{1:Vallee97} J.~P.~Vall\'ee: Fund.~Cosm.~Phys. {\bf 19}, 1 (1997)

\bibitem{1:kolmogorov} A.~M.~Obukhov: Dokl.~Akad.~Nauk~SSSR {\bf 32},
22 (1941)\\
A.~N.~Kolmogorov: Dokl.~Akad.~Nauk~SSSR
{\bf 30}, 299 (1941); ibid. {\bf 31}, 538 (1941); ibid.
{\bf 32}, 19 (1941)\\
W.~Heisenberg: Z.~Physik {\bf 124}, 628 (1948)\\
R.~Z.~Sagdeev: Rev.~Mod.~Phys. {\bf  51}, 1 (1979)\\
M.~L.~Goldstein, D.~A.~Roberts, W.~H.~Matthaeus:
Ann.~Rev.~Astron.~Astroph. {\bf 33}, 283 (1995)

\bibitem{1:Krause98} F.~Krause, R.~Beck: Astron.~Astrophys. {\bf 335},
789 (1998)
% symmetry of magnetic fields in spirals, seems to be point inwards along
%  spiral arms

\bibitem{1:Kulsrud99} R.~M.~Kulsrud: Ann.~Rev.~Astron.~Astrophys. {\bf 37},
37 (1999)
%  A Critical Review of Galactic Dynamos

\bibitem{1:GR2001} D.~Grasso, H.~R.~Rubinstein: Phys.~Rep. {\bf 348},
163 (2001)

\bibitem{1:Nath94} B.~B.~Nath, P.~L.~Biermann: Month.~Not.~R.~Astron.~Soc.
{\bf 267}, 447 (1994)
%    title = "Cosmic ray ionization of the interstellar medium",

\bibitem{1:RKB98} D.~Ryu, H.~Kang, P.~L.~Biermann: Astron.~Astrophys.
{\bf 335}, 19 (1998)
%  Cosmic magnetic field in large scale filament and sheets, 

\bibitem{1:Piran99b} T.~Piran: Phys. Rep. {\bf 314}, 575 (1999)
%  Title:  Gamma-ray bursts and the fireball model.
%

\bibitem{1:SM96} D.~B.~Sanders, I.~F.~Mirabel:
Ann.~Rev.~Astron.~Astrophys. {\bf 34}, 749 (1996)
%  review on ultraluminous infrared galaxies

\bibitem{1:LM2000} J.~G.~Learned, K.~Mannheim: Ann.~Rev.~Nucl.~Part.~Sci.
{\bf 50}, 679 (2000)
%  Title:  High energy neutrino astrophysics
%

\bibitem{1:gut} see, e.g., S.~Weinberg: {\it The Quantum Theory of
Fields Vol 2: Modern Applications}, Cambridge University Press
(Cambridge, 1996)

\bibitem{1:evans} see, e.g., W.~Evans, F.~Ferrer, S.~Sarkar:
astro-ph/0103085; to appear in Astropart.~Phys.

\bibitem{1:kibble} T.~W.~B.~Kibble: J.~Phys. {\bf A9}, 1387
(1976)\\
for a review see A.~Vilenkin: Phys.~Rep. {\bf 121}, 263 (1985)

\bibitem{1:kt} see, e.g., E.~W.~Kolb, M.~S.~Turner:
{\it The Early Universe} (Addison-Wesley, Redwood City, California, 1990)

\bibitem{1:kuz-tak} for a brief review see V.~Kuzmin, I.~Tkachev:
Phys.~Rep. {\bf 320}, 199 (1999)

\bibitem{1:bere-kachel} V.~Berezinsky, M.~Kachelriess: Phys.~Rev.~D
{\bf 63} 034007 (2001)

\bibitem{1:wkwb} S.~D.~Wick, T.~W.~Kephart, T.~J.~Weiler, P.~L.~Biermann:
astro-ph/0001233

\bibitem{1:barrau} A.~Barrau: Astropart.~Phys. {\bf 12}, 269 (2000)

\bibitem{1:Zburst} T.~J.~Weiler: Phys.~Rev.~Lett. {\bf  49}, 234 (1982)\\
T.~J.~Weiler: Astrophys.~J. {\bf 285}, 495 (1984)\\
T.~J.~Weiler: Astropart.~Phys. {\bf 11}, 303 (1999)\\
D.~Fargion, B.~Mele, A.~Salis: Astrophys.~J. {\bf 517}, 725 (1999)\\
G.~Gelmini, A.~Kusenko: Phys.~Rev.~Lett. {\bf 82}, 5202 (1999);
ibid. {\bf 84}, 1378 (2000)

\bibitem{1:kascade} see, e.g., K.-H.~Kampert, et al.: astro-ph/0102266

\bibitem{1:antimatter} see, e.g., WiZard/CAPRICE Collaboration:
astro-ph/0103513\\
T.~Maeno, et al., BESS Collaboration: Astropart.~Phys. {\bf 16}, 121 (2001)\\
M.~A.~Huang: astro-ph/0104229

\bibitem{1:Shapiro99} M.~M.~Shapiro: in {\it LiBeB, Cosmic rays, and related
X- and $\gamma$-rays}, ASP conf. No. 171, Eds. R. Ramaty \etal, p. 139 -
145 (1999)
%  Title:  How cosmic rays get started

\bibitem{1:Schmitt94}  J.~H.~M.~M. Schmitt, B.~M.~Haisch, J.~J.~Drake:
Science {\bf 265}, 1420+ (1994)
%  title = "A Spectroscopic Measurement of the Coronal Density of
%  Procyon",

\bibitem{1:Drake95} J.~J.~Drake, et al.: Science {\bf 267}, 1470+ (1995)
%  title = "The Elemental Composition of the Corona of Procyon - Evidence
%  for the Absence of the FIP Effect",

\bibitem{1:JPMeyer97} J.-P.~Meyer, L.~O'C~Drury, D.~C.~Ellison: Astrophys.~J. 
{\bf 487}, 182 (1997)
%  Title:  Galactic Cosmic Rays from Supernova Remnants. I. A
%  Cosmic-Ray Composition Controlled by  Volatility and Mass-to-Charge Ratio

\bibitem{1:Rickett90} B.~J.~Rickett: Ann.~Rev.~Astron.~Astroph.
{\bf 28}, 561 (1990)
%

\bibitem{1:BZ34} W.~Baade, F.~Zwicky: Proc.~Nat.~Acad.~Science {\bf 20},
no. {\bf 5}, 259 (1934)

\bibitem{1:borione} A.~Borione, et al.: Astrophys.~J. {\bf 493}, 175 (1998)

%\bibitem{1:GamowCR} P.~L.~Biermann: Space~Science~Rev. {\bf 74}, 385 (1995)\\
%P.~L.~Biermann: Phys. Rev.~D {\bf 51}, 3450 (1995)
%  The origin of cosmic rays, Peter L. Biermann; invited 

\bibitem{1:SeoPt94} E.-S.~Seo, V.~S.~Ptuskin: Astrophys.~J. {\bf 431},
705 (1994)
 title = "Stochastic reacceleration of cosmic rays in the
  interstellar medium",

\bibitem{1:TucsonCR} P.~L.~Biermann: in {\it Cosmic winds and the Heliosphere},
Eds. J. R. Jokipii, et al., Univ. of Arizona press, p. 887 - 957 (1997),
astro-ph/9501030
%  Title: Supernova blast waves and pre-supernova winds: Their cosmic
%  ray contribution

\bibitem{1:CRIX} P.~L.~Biermann, N.~Langer, E.~Seo, T.~Stanev:
Astron.~Astrophys. {\bf 369}, 269 (2001)
%  title = "Cosmic rays IX. Interactions and transport of cosmic rays in
%  the Galaxy"

\bibitem{1:moskalenko} A.~W.~Strong, I.~V.~Moskalenko: astro-ph/0101068,
invited talk at COSPAR2000 Symposium, E1.3: Origin and Acceleration of Cosmic
Rays, ed. M.H. Israel , to appear in Adv. Sp. Res. Vol. {\bf 27} No. 4\\
I.~V.~Moskalenko, A.~W.~Strong: Astrophys.~Sp.~Sci. {\bf 272}, 247 (2000),
and references therein

\bibitem{1:CASAMIA} M.~A.~K.~Glasmacher, et al.: Astropart.~Phys. {\bf 10}, 
291 (1999); ibid. {\bf 12}, 1 (1999)

\bibitem{1:kangetal} H.~Kang, D.~Ryu, T.~W.~Jones: Astropart.~Phys.
{\bf 456}, 422 (1996)\\
H.~Kang, J.~P.~Rachen, P.~L.~Biermann: \MNRAS {\bf 286}, 257 (1997)

\bibitem{1:Hillas84} A.~M.~Hillas: Ann.~Rev.~Astron.~Astroph.
{\bf 22}, 425 (1984)

\bibitem{1:BS87} P.~L.~Biermann, P.~A.~Strittmatter: Astrophys.~J.
{\bf 322}, 643 (1987)
%
\bibitem{1:bs} for a detailed review see P.~Bhattacharjee, G.~Sigl:
Phys.~Rep. {\bf 327}, 109 (2000)

\bibitem{1:linvar} S.~Coleman, S.~L.~Glashow: Phys.~Rev.~D {\bf 59}, 116008
(1999)\\
G.~Amelino-Camelia, J.~Ellis, N.~E.~Mavromatos, D.~V.~Nanopoulos,
S.~Sarkar: Nature {\bf 393}, 763 (1998)\\
T.~Kifune, Astrophys.~J.~Lett. {\bf 518}, L21 (1999)\\
R.~Aloisio, P.~Blasi, P.~L.~Ghia, A.~F.~Grillo, Phys.~Rev.~D {\bf 62},
053010 (2000)

\bibitem{1:ph} H.~Meyer, R.~Protheroe: Phys.~Lett. {\bf B493}, 1 (2000)

\bibitem{1:acp} G.~Amelino-Camelia, T.~Piran: Phys.~Lett. {\bf B497}, 265
(2001)

\end{thebibliography}
\end{document}